\documentclass[manuscript=article, layout=onecolumn]{achemso}
\setkeys{acs}{articletitle = true}

\usepackage[T1]{fontenc}
\usepackage[utf8]{inputenc}
\usepackage{graphicx}
\usepackage{amsmath}
\usepackage{color}
\usepackage{comment}
\usepackage{hyperref}
\usepackage{multirow}
\usepackage{subfig}
\usepackage{bm}
\usepackage{braket}
\usepackage{siunitx}
\usepackage[normalem]{ulem}   
\usepackage[version=4]{mhchem}
\usepackage{arydshln}

\definecolor{purple}{rgb}{0.5,0,0.5}
\newcommand{\emon}[1]{{\color{purple} #1}}
\definecolor{ruby}{rgb}{0.88, 0.07, 0.37}

\DeclareSIUnit{\kcal}{\kilo\cal}

\SectionNumbersOn 

\title{
Why Projection-Based DMRG-in-DFT Cannot Be Exact, Even with the
Exact Exchange-Correlation Functional}

\author{Enzo Monino}
\affiliation{J. Heyrovsk\'{y} Institute of Physical Chemistry, Academy of Sciences of the Czech \mbox{Republic, v.v.i.}, Dolej\v{s}kova 3, 18223 Prague 8, Czech Republic}
\altaffiliation{Contributed equally.}

\author{Daria Drwal}
\affiliation{Institute of Physics, Lodz University of Technology, \mbox{ul.\ Wolczanska 217/221, 93-005 Lodz, Poland}}
\altaffiliation{Contributed equally.}

\author{Micha{\l} Hapka}
\affiliation{University of Warsaw, Faculty of Chemistry, ul.\ L.\ Pasteura 1, 02-093 Warsaw, Poland}

\author{Libor Veis}
\email{libor.veis@jh-inst.cas.cz}
\affiliation{J. Heyrovsk\'{y} Institute of Physical Chemistry, Academy of Sciences of the Czech \mbox{Republic, v.v.i.}, Dolej\v{s}kova 3, 18223 Prague 8, Czech Republic}

\author{Katarzyna Pernal}
\affiliation{Institute of Physics, Lodz University of Technology, \mbox{ul.\ Wolczanska 217/221, 93-005 Lodz, Poland}}
\email{pernalk@gmail.com}

\keywords{exact wave function in density functional theory embedding, strong correlation, density matrix renormalization group, fractional spin error}

\begin{document}

\begin{abstract}



We establish the theoretical foundations for embedding a correlated wave function in an environment formed by Kohn–Sham orbitals. We show that introducing an approximation which equates two, in principle distinct, kinetic-energy functionals yields an embedding functional identical to the projection-based wavefunction-in-DFT formulation of Miller and co-workers. We demonstrate that this functional is inherently nonvariational: its minimum is not guaranteed to coincide with the exact ground-state energy and remains bounded from above by it. Building on this formal framework, we analyze the dominant sources of error in projection-based DMRG-in-DFT embedding with approximate exchange–correlation (xc) functionals. Using molecules with dissociating covalent bonds as a diagnostic example, we demonstrate that the primary source of error is the nonadditive exchange–correlation energy describing the nonclassical coupling between the active subsystem and its environment. Eliminating the fractional-spin error by employing a pair-density xc functional (PDFT) instead of a semilocal GGA does not remedy this deficiency, because the inaccuracy stems from self-interaction effects at the subsystem–environment interface.

\end{abstract}

\section{Introduction}

Strong correlation is prevalent across chemistry, influencing
bond dissociation, open-shell and excited-state electronic structures, and catalytic reactivity.~\cite{lyakh2012multireference, Szalay2011} Accurate description of strongly correlated systems remains a formidable challenge from an electronic structure perspective. Although full configuration interaction (FCI) method provides exact results within a given single-particle basis, its computational cost is prohibitive. Consequently, only small systems can be treated at this level of theory. This limitation has motivated the development of approximate wave function (WF) methods that are systematically improvable toward the FCI limit. For weakly correlated systems, the coupled-cluster (CC) method~\cite{Bartlett2007} is the method of choice and probably the most popular one. Strongly correlated systems are more suitably described within a complete active space (CAS) framework,~\cite{roos1987complete} most notably by the complete active space self-consistent field (CASSCF) method.~\cite{Roos1980} CASSCF combines a FCI treatment of a chosen set of active orbitals with orbital optimization. However, its applicability is limited by the exponential scaling of the FCI step, which restricts the size of the CAS to roughly 20 orbitals.~\cite{Vogiatzis2017} A variety of approximate FCI solvers have been developed to overcome this bottleneck, including FCI quantum Monte Carlo (FCIQMC) \cite{Booth2009}, selected configuration interaction methods such as heat-bath CI (HCI) \cite{Holmes2016}, and many-body expansion (MBE) approaches.~\cite{Eriksen2018, Eriksen2019} Among these, the density matrix renormalization group (DMRG) method~\cite{White1992} has emerged as one of the most powerful and widely used approaches for treating strong electron correlation.~\cite{chan_review,Szalay2015,yanai_review,reiher_perspective, Golub2021}

On a different path, density functional theory (DFT) is, in principle, exact, and its most common approximate formulation,  Kohn-Sham (KS) DFT, provides a computationally efficient alternative to WF methods. As a result, much larger systems can be described at this level. The main limitation of KS-DFT is the use of approximate exchange-correlation (xc) functionals, which have a single-reference character and are therefore unreliable for strongly correlated systems.~\cite{burke2012}

Combining the capabilities and accuracy of WF-based approaches with the computational lightness of KS-DFT through quantum embedding methods is a compelling idea. In general, embedding frameworks~\cite{Jones2020} seek to go beyond the system size limits by exploiting the relative locality of electronic interactions and separating the full system into an active region and the environment. The active region is treated with a high-level quantum mechanical method and is embedded in its environment, described with a lower-level method. Prominent examples include quantum mechanics/molecular mechanics (QM/MM) \cite{Senn2009}, density matrix embedding theory (DMET) \cite{Knizia2012, Wouters2016}, Green's function embedding \cite{Lin2011, Lan2015}, and DFT embedding \cite{Jacob2014, jacob2024subsystem,Lee2019} schemes. Complementary embedding strategies, such as the subsystem embedding sub-algebra formalism \cite{Kowalski2018}, underpin recent developments in active-space coupled-cluster downfolding techniques \cite{Bauman2022}.

This work focuses on systems that contain a strongly correlated fragment, in particular those involving dissociating chemical bonds. In such cases, it is desirable to combine WF-based methods capable of accurately describing the strongly correlated active region, here, the DMRG method, with DFT to model the surrounding environment. 

The theoretical foundations of DFT embedding were established in the pioneering work of Cortona \cite{Cortona1991}, and later by Wesolowski and Warshel \cite{Wesolowski1993}. The latter formulated the frozen density embedding (FDE) framework \cite{Wesolowski2008,wesolowski2015frozen}, in which the total electron density is partitioned into an active subsystem treated variationally and an environment represented by a frozen density, coupled through an embedding potential containing nonadditive exchange-correlation and kinetic energy terms. While the nonadditive exchange-correlation contribution can be treated within standard density functionals, approximations to the nonadditive kinetic energy potential (NAKP), required to enforce Pauli exclusion between subsystems, generally fail for systems with significant density overlap, thereby restricting the applicability of FDE to weakly interacting or noncovalent systems. Although formally exact approaches for evaluating the NAKP have been proposed, they entail substantial additional computational complexity and introduce further theoretical and numerical challenges.\cite{goodpaster2010exact}

The projection-based DFT (PB-DFT) formalism \cite{Lee2019} was introduced to circumvent the limitations of FDE by removing the NAKP contributions, enforcing the orthogonality between occupied orbitals of the active and environment parts. This orthogonality condition was initially enforced by the use of a level-shift projection operator \cite{Manby2012}, but parameter-free formulations also exist~\cite{Khait2012,Graham2020}. The PB-DFT framework is an exact embedding scheme, i.e., if both parts are treated at the DFT level with the same xc functional, then the PB-DFT energy is equal to the KS-DFT energy of the full system. In practice, however, PB-DFT is primarily designed for WF-in-DFT embedding, wherein the active subsystem is described using a high-level wavefunction method while the surrounding environment is treated at the DFT level. Until now, it has remained unclear whether PB-WF-in-DFT embedding is an exact theory, i.e.\ whether it can formally recover the exact ground-state energy if the exact exchange-correlation functional were available.

In our previous work, we introduced the projection-based DMRG-in-DFT framework as an efficient embedding approach for systems with strongly correlated fragments \cite{Beran2023}, and subsequently proposed a correction of nonadditive xc errors using the multireference adiabatic connection method \cite{Monino2026}. In the present article, we establish the theoretical foundations of projection-based wavefunction-in-DFT embedding by constructing an exact energy functional for a wavefunction embedded in a Kohn–Sham environment. We demonstrate that, similarly to orbital-free embedding, this formulation involves an unknown kinetic-energy functional. We further show that the standard formulation of projection-based WF-in-DFT embedding\cite{Manby2012} renders this functional nonvariational, implying that the latter is not formally exact even when the exact xc functional is employed. Finally, we analyze the dominant sources of error in WF-in-DFT embedding for strongly correlated systems and demonstrate that the fractional-spin error of approximate xc functionals is not the principal source of inaccuracies in dissociating systems. In fact,  the use of a fractional-spin-error-free approach, such as pair-density functional theory (PDFT) \cite{pdft}, may lead to larger errors than conventional density functionals.

This work is organized as follows. In Section~\ref{section_theory}, we present a pedagogical and detailed comparison between the orbital-free embedding scheme and embedding in a Kohn-Sham orbital environment. In this section, we also present our theoretical results, namely the existence of an exact energy functional for a multi-determinantal wave function embedded in a Kohn-Sham orbital environment and its explicit dependence on the kinetic-energy functional. In the subsequent sections, we focus on analyzing the sources of error in the nonadditive xc energy contributions. Specifically, Section~\ref{section_comp_details} describes the computational details of the benchmark calculations, while Section~\ref{section_results} discusses the obtained results. Our conclusions are presented in Section~\ref{section_conclusions}.

\section{Theory}
\label{section_theory}
The aim of this section is to introduce a general and exact framework for embedding a wave function in an environment described by a subset of known Kohn–Sham orbitals. We compare this formalism with its analogue in orbital-free environment embedding theory, as presented in Ref.~\citenum{Wesolowski2008}. The central result  is that the projection-based WF-in-DFT functional introduced in Ref.~\citenum{goodpaster2014accurate} cannot, in general, yield the exact ground-state energy, even if it were employed with the exact exchange–correlation functional.

For clarity, we first introduce the notation, definitions, and several useful relations. We then discuss Kohn–Sham embedding in an orbital environment and compare this formulation with Kohn–Sham system embedding in an orbital-free environment. Finally, we present the main results, demonstrating how a wave function can be rigorously and exactly embedded in a Kohn–Sham orbital environment and identifying the assumptions needed to obtain the projection-based WF-in-DFT functional from the exact formulation.

\subsection{Definitions and fundamental relations}
Recall that for a given electron density $\rho$ integrating to a number of electrons $N$, a noninteracting kinetic energy density functional, $T_{s}[\rho]$, is defined via a constrained minimization of the kinetic energy with respect to $N$ orthonormal spinorbitals  that reproduce $\rho$
\begin{equation}
T_s[\rho]=\min_{\sum_{i=1}^{N}|\varphi_{i}|^2=\rho}\sum_{i=1}^N \left\langle \varphi_{i}|\hat{t}|\varphi_{i}\right\rangle
\end{equation}
Introduce a kinetic energy orbital functional%
\begin{equation}
T[\left\{  \varphi_{i}\right\}  _{N}]=\sum_{i=1}^{N}\left\langle \varphi
_{i}|\hat{t}|\varphi_{i}\right\rangle \ \ \ \label{T}%
\end{equation}
where $\left\{  \varphi_{i}\right\}  _{N}$ denotes a set of $N$ orthonormal
spinorbitals. 
Notice the following inequality between the $T$ and $T_s$ functionals
\begin{eqnarray}
\forall_{\left\{  \varphi_{i}\right\}  _{N}}\ \ \ 
T[\left\{  \varphi_{i}\right\}  _{N}]\geq T_{s}[\rho]\ \ \ \label{TTs}%
\end{eqnarray}
where
\begin{equation}
\rho(\mathbf{x})=\sum_{i=1}^{N}\left\vert \varphi_{i}(\mathbf{x})\right\vert
^{2}\ \ \ \label{rho_phi}%
\end{equation}
and $\mathbf{x}$ denotes combined spatial and spin coordinates.

Consider an $N$-electron interacting system in an external potential $V_{ext}$
and let a ground state electron density $\rho_{0}$ be non-interacting pure
state v-representable. A KS functional $E^{\text{KS}}$ is defined for a set of
$N$ orthonormal spinorbitals%
\begin{equation}
E^{\text{KS}}[\left\{ \varphi_{i} \right\}_{N}] = T[\left\{  \varphi_{i}\right\}_{N}] + V_{ext}[\rho] + E_{Hxc}[\rho]\ \ \ \label{EKS}%
\end{equation}
where the Hartree-exchange-correlation density functional, $E_{Hxc}$, is
defined through a Levy constraint search formula
\begin{equation}
E_{Hxc}[\rho] = \min_{\Psi_{N}\rightarrow\rho} \left\langle \Psi_{N}|\hat{T}
 + \hat{V}_{ee}|\Psi_{N} \right\rangle - T_{s}[\rho]\ \ \ \label{EHxc}
\end{equation}
$\Psi_{N}$ indicates an $N$-electron antisymmetric function. A\ minimum of the
KS\ functional is equal to a ground state energy $E_{0}$
\begin{equation}
E_{0} = \min_{\left\{  \varphi_{i}\right\}_{N}}\ E^{\text{KS}}[\left\{
\varphi_{i}\right\}  _{N}]=T[\left\{  \varphi_{i}^{\text{KS}}\right\}_{N}] + V_{ext}[\rho_{0}] + E_{Hxc}[\rho_{0}]\ \ \ \ \label{E0}%
\end{equation}
and the minimizing orbitals are the Kohn-Sham orbitals $\left\{ \varphi_{i}^{\text{KS}}\right\}  $, which yield an exact ground state electron density $\rho_{0}$
\begin{equation}
\rho_{0}(\mathbf{x})=\sum_{i=1}^{N}\left\vert \varphi_{i}^{\text{KS}%
}(\mathbf{x})\right\vert ^{2}\ \ \
\end{equation}

A Hohenberg-Kohn (HK) density functional, $E^{\text{HK}}[\rho]$, for noninteracting
v-representable densities can be written in terms of $T_{s}$ and
 $E_{Hxc}$ functionals, namely%
\begin{equation}
E^{\text{HK}}[\rho]=T_{s}[\rho]+V_{ext}[\rho]+E_{Hxc}[\rho]\ \ \ \label{HK}%
\end{equation}
and it coincides at the minimum with the ground state energy%
\begin{equation}
E_{0}=\min_{\rho}\text{ }E^{\text{HK}}[\rho]=E^{\text{HK}}[\rho_{0}]\ \ \
\end{equation}
Notice a general relation, valid for electron density and orbitals related by
Eq.~(\ref{rho_phi}), between the HK\ and KS\ functionals
\begin{equation}
\forall_{\left\{  \varphi_{i}\right\}  _{N}} \ \ \ E^{\text{KS}}[\left\{  \varphi_{i}\right\}  _{N}]-E^{\text{HK}}[\rho]\geq0
\end{equation}
resulting immediately from the inequality in Eq.~\eqref{TTs}.

\subsection{Setting the problem}

Assume that for an $N$-electron system, a subset of $N_{B}$ Kohn-Sham
orbitals, denoted as $\left\{  \varphi_{i}^{\text{KS}}\right\}_{N_{B}}$, is known.
A\ corresponding electron density $\rho_{B}$
\begin{equation}
\rho_{B}(\mathbf{x})=\sum_{i}^{N_{B}}\left\vert \varphi_{i}^{\rm KS}
(\mathbf{x})\right\vert ^{2}\ \ \ \label{rho_B}%
\end{equation}
forms an ``environment'' ($B$) for the ``active'' ($A$) part of the system comprising $N_A=N-N_B$ electrons and described by an unknown electron density
$\rho_{A}$, such that
\begin{equation}
\rho_{A}(\mathbf{r})=\rho_{0}(\mathbf{r})-\rho_{B}(\mathbf{r}%
)\ \ \label{rho_0}
\end{equation}

The problem  of interest is the following: given $\left\{ \varphi_i^{\text{KS}}\right\}_{N_B}$, describe an active system by a quantum descriptor $Q$ and formulate an energy functional of Q  over the proper domain, such that its minimum equals the ground state energy $E_{0}$. Our target will be constructing a functional when $Q$ is chosen to be a correlated $N_{A}$-electron wavefunction. First, however, we consider a case when $Q$ is a set of $N_{A}$ spinorbitals.

\subsection{KS-DFT-in-KS-DFT: embedding orbitals in KS orbital-environment}

Before moving to wavefunction embedding, let us consider a  problem of finding a functional when $Q$ is chosen as a set of $N_{A}$ orthonormal orbitals describing an active system. 
For a given set of environment KS orbitals $\left\{ \varphi_{i}^{\rm KS} \right\}_{N_B}$, define a functional $E^{\text{DFT-in-DFT}}$, depending on $N_A$ orthonormal orbitals $\left\{ \varphi_{i}\right\}_{N_A}$ reading
\begin{gather}
E^{\text{DFT-in-DFT}}[\left\{  \varphi_{i}\right\}_{N_A};\left\{
\varphi_{i}^{\rm KS}\right\}_{N_B}] = E^{\text{KS}}[\left\{  \varphi
_{i}\right\}_{N_A} \cup \left\{ \varphi_{i}^{\rm KS}\right\}_{N_B}] \nonumber\\
= T[\left\{  \varphi_{i}\right\}_{N_A}] + T[\left\{ \varphi_{i}^{\rm KS} \right\}_{N_B}] + V_{ext}[\rho_{A}+\rho_{B}] + E_{Hxc}[\rho_A + \rho_B] \ \ \ \label{EAB}
\end{gather}
where $\rho_{B}$ is fixed, see Eq.~\eqref{rho_B}, and
\begin{equation}
\rho_{A}(\mathbf{x})=\sum_{i}^{N_{A}}\left\vert \varphi_{i}(\mathbf{x}%
)\right\vert ^{2} \label{rho_A}%
\end{equation}
To assure that the functional is bounded by the exact energy $E_{0}$, spinorbitals
in the set $\left\{  \varphi_{i}\right\}_{N_A}$ must be constrained to be
orthogonal to those in  $\left\{  \varphi_{i}^{\rm KS}\right\}_{N_B}$, which
we denote as $\left\{  \varphi_{i}\right\}_{N_A}\perp\left\{  \varphi_{i}^{\rm KS}\right\}  _{N_B}$, i.e.
\begin{equation}
\left\{  \varphi_{i}\right\}_{N_A} \perp \left\{ \varphi_{i}^{\rm KS}\right\}_{N_B}\ \ \ \Longrightarrow\ \ \ \forall_{\substack{\varphi_{j}\in\left\{
\varphi_{i}\right\}_{N_A} \\
\varphi_{k} \in \left\{ \varphi_{i}^{\rm KS}\right\}_{N_B}}}\ \ \ \left\langle \varphi_{j}|\varphi_{k}
\right\rangle = 0 \label{ortho}%
\end{equation}
By confronting the inequality in Eq.~\eqref{TTs}, it is evident that the kinetic energy in Eq.~\eqref{EAB} is bounded from below by a noninteracting kinetic energy functional of combined densities $\rho_{A}+\rho_{B}$
\begin{equation}
\forall_{\left\{  \varphi_{i}\right\}_{N_A} \perp \left\{ \varphi_{i}^{\rm KS}\right\}_{N_B} }\ \ \ T[\left\{  \varphi_{i}\right\}_{N_A}] + T[\left\{  \varphi_{i}^{\rm KS}\right\}_{N_B}] \geq T_{s}[\rho_{A}+\rho_{B}]\ \ \ \label{ineq_2}
\end{equation}
Notice that without the assumption of the orthogonality of orbitals, condition
in Eq.~\eqref{ortho}, the above inequality would not be generally valid. Eqs.~\eqref{ineq_2} and (\ref{HK}) imply that
\begin{equation}
\forall_{\left\{  \varphi_{i}\right\}  _{N_{A}}\perp\left\{  \varphi_{i}^{\rm KS}\right\}  _{N_B}}\ \ \ E^{\text{DFT-in-DFT}}[\left\{  \varphi_{i}\right\}_{N_A};\left\{  \varphi_{i}^{\rm KS}\right\}_{N_B}]\geq
E^{\text{HK}}[\rho_{A}+\rho_{B}]\geq E_{0}\ \ \ \label{rel2}
\end{equation}

Therefore, minimization of $E^{\text{DFT-in-DFT}}$ in a domain restricted by
the orthogonality condition, Eq.~\eqref{ortho}, leads to a ground state energy
\begin{equation}
\min_{\left\{  \varphi_{i}\right\}_{N_A} \perp \left\{  \varphi_{i}^{\rm KS}\right\}_{N_B}}\ \ E^{\text{DFT-in-DFT}}[\left\{  \varphi_{i}\right\}_{N_A};\left\{  \varphi_{i}^{\rm KS}\right\} _{N_B}] = E_{0}%
\end{equation}
The minimizing orbitals are the KS\ orbitals complementary to $\left\{
\varphi_{i}^{\rm KS}\right\}_{N_B}$, namely
\begin{equation}
\arg\min_{\left\{ \varphi_{i}\right\}_{N_A} \perp \left\{  \varphi_{i}^{\rm KS}\right\}_{N_B}}\ \ E^{\text{DFT-in-DFT}}[\left\{  \varphi_{i}\right\}_{N_A}; \left\{  \varphi_{i}^{\rm KS}\right\}_{N_B}]=\left\{
\varphi_{i}^{\rm KS}\right\}_{N_A}\ \ \ \label{rel5}%
\end{equation}

It is instructive to compare functionals for orbital- and
orbital-free-environment embedding. The latter is a central object in the
frozen density embedding theory (FDET)\cite{Wesolowski2008}, where it is assumed that only the
electron density is given for the environment. Its definition reads 
\begin{equation}
E^{\text{FDET}}[\left\{  \varphi_{i}\right\}  _{N_{A}},\rho_{B}]=T[\left\{
\varphi_{i}\right\}  _{N_{A}}]+T_{s}[\rho_{B}]+T_{s}^{\text{nadd}}[\rho
_{A},\rho_{B}]+V_{ext}[\rho_{A}+\rho_{B}]+E_{Hxc}[\rho_{A}+\rho_{B}]
\end{equation}
where $T_{s}^{\text{nadd}}$ denotes a nonadditive kinetic energy functional,
\begin{equation}
T_{s}^{\text{nadd}}[\rho_{A},\rho_{B}]=T_{s}[\rho_{A}+\rho_{B}]-T_{s}[\rho
_{A}]-T_{s}[\rho_{B}]
\end{equation}
Due to the presence of $T_{s}^{\text{nadd}}$ in the functional 
$E^{\text{FDET}}[\left\{  \varphi_{i}\right\}  _{N_{A}},\rho_{B}]$, by exploiting
Eq.~(\ref{TTs}) it is straightforward to conclude that it is bounded from below
by the HK\ energy $E^{\text{HK}}[\rho_{A}+\rho_{B}]$ for any set $\left\{
\varphi_{i}\right\}  _{N_{A}}$
\begin{equation}
E^{\text{FDET}}[\left\{  \varphi_{i}\right\}  _{N_{A}},\rho_{B}]\geq E^{\text{HK}%
}[\rho_{A}+\rho_{B}]\ \ \ 
\end{equation}
In general, minimization of the functional $E^{\text{FDET}}$ leads to a ground state energy $E_{0}$ but the minimizing orbitals need not coincide with the KS\ orbitals of the composite system. The orbital-environment embedding functional, see Eq.~\eqref{EAB}, does not involve the $T_{s}[\rho]$ functional but $E^{\text{DFT-in-DFT}}$ minimization is constrained by the orthogonality condition shown Eq.~\eqref{ortho}.

\subsection{WF-in-KS-DFT: embedding wavefunction in KS orbital-environment}

\subsubsection{Construction of the exact functional}
In this section, we assume that the active subsystem, comprising $N_{A}$
electrons, is described via a correlated wavefunction $\Psi_{N_{A}}$ and we
seek a pertinent functional, which at the minimum yields a ground state energy
$E_{0}$. To achieve this goal, begin by adding and subtracting a HK functional
of density $\rho_{A}$ in the functional $E^{\text{DFT-in-DFT}}$,
Eq.~(\ref{EAB}),
\begin{align}
E^{\text{DFT-in-DFT}}[\left\{  \varphi_{i}\right\}_{N_A};\left\{
\varphi_{i}^{\rm KS}\right\}_{N_B}] & =\min_{\Psi_{N_{A}}\rightarrow\rho_{A}} \left\langle \Psi_{N_{A}}|\hat{T}+\hat{V}_{ee}+\hat{V}_{ext}|\Psi_{N_{A}%
}\right\rangle +E^{\text{KS}}[\left\{  \varphi_{i}\right\}  _{N_{A}}%
\cup\left\{  \varphi_{i}^{\rm KS}\right\}  _{N_{B}}]\nonumber\\
&  -E^{\text{HK}}[\rho_{A}]\ \ \
\end{align}
where $\rho_{A}$ is given by the orbitals $\left\{  \varphi_{i}\right\}_{N_A}$, see Eq.~\eqref{rho_A}. To define a functional of $\Psi_{N_A}$, the middle term in the above expression is turned into a functional of the density $\rho_A$ corresponding to a wavefunction $\Psi_{N_A}$
\begin{equation}
\rho_{A}(\mathbf{r})\equiv \rho_{A}[\Psi_{N_{A}}](\mathbf{r})=\left\langle
\Psi_{N_{A}}|\sum_{i}^{N_{A}}\delta(\mathbf{r}_{i}\mathbf{-r})|\Psi_{N_{A}%
}\right\rangle
\end{equation}
and the functional is defined as%
\begin{align}
E^{\text{WF-in-DFT}}[\Psi_{N_{A}};\left\{  \varphi_{i}^{\rm KS}\right\}_{N_B}] &= \left\langle \Psi_{N_A}|\hat{T}+\hat{V}_{ee}+\hat{V}_{ext}
|\Psi_{N_A}\right\rangle + \min_{\substack{\left\{  \varphi_{i}\right\}_{N_A}\rightarrow\rho_{A} \\ 
\left\{  \varphi_{i}\right\}_{N_A} \perp \left\{  \varphi_{i}^{\rm KS}\right\}_{N_B}}} E^{\text{KS}}[\left\{ \varphi_{i}\right\}_{N_A}\cup\left\{  \varphi_{i}^{\rm KS}\right\}_{N_B}] \nonumber \\
& - E^{\text{HK}}[\rho_{A}]\label{exact}
\end{align}
A domain of this functional is restricted to wavefunctions $\Psi_{N_A}$ belonging to a set $\Omega_{A}$ constructed for a given set $\left\{ \varphi_{i}^{\rm KS} \right\} _{N_B}$ as
\begin{equation}
\Omega_{A}[\left\{ \varphi_{i}^{\rm KS} \right\} _{N_B}]=\left\{  \Psi_{N_A}\  :\  \ \exists\ \left\{  \varphi_{i}\right\}  _{N_A}\ \ \ \sum_{i}^{N_A}\left\vert \varphi_{i}(\mathbf{x})\right\vert ^{2}=\rho_A
[\Psi_{N_A}](\mathbf{x}) \ \wedge  \ \left\{  \varphi_{i}\right\}
_{N_A}\perp\left\{  \varphi_{i}^{\rm KS}\right\}_{N_B}\right\}
\ \ \ \label{OmA}
\end{equation}
Thus, $\Omega_{A}$ comprises $N_{A}$-electron wavefunctions such that the corresponding electron
density $\rho_A$ is representable by a set of $N_{A}$ orbitals
orthogonal to those in the set $\left\{ \varphi_{i}^{\rm KS}\right\}_{N_B}$.
This requirement assures that the middle term in Eq.~\eqref{exact} is well defined. Moreover, since $\left\langle \Psi_{N_A}|\hat
{T}+\hat{V}_{ee}+\hat{V}_{ext}|\Psi_{N_A}\right\rangle \geq E^{\text{HK}}[\rho_{A}]$, then
\begin{equation}
\forall_{\Psi_{N_{A}}\in\Omega_{A}}\ \ \ E^{\text{WF-in-DFT}}[\Psi_{N_A};\left\{  \varphi_{i}^{\rm KS}\right\}_{N_B}] \geq \min_{\substack{\left\{
\varphi_{i}\right\}_{N_A} \rightarrow \rho_{A}[\Psi_{N_A}]\\\left\{  \varphi_{i}\right\}_{N_A} \perp \left\{  \varphi_{i}^{\rm KS}\right\}_{N_B}}} E^{\text{KS}}[\left\{  \varphi_{i}\right\}_{N_A} \cup \left\{  \varphi_i^{\rm KS}\right\}_{N_B}] \geq E^{\text{HK}}[\rho_{A}+\rho_{B}]\
\end{equation}
where the inequality (\ref{ineq_2}) has been employed. Consequently,
minimization of the functional $E^{\text{WF-in-DFT}}$ over $\Omega_{A}$
yields a ground state energy,%
\begin{equation}
E_{0}=\min_{\Psi_{N_{A}}\in\Omega_{A}}\ E^{\text{WF-in-DFT}}[\Psi_{N_A};\left\{  \varphi_{i}^{\rm KS}\right\}  _{N_{B}}]\ \ \ \label{E0min}%
\end{equation}
as desired.
Notice that a minimizer, $\ \Psi_{N_{A}}^{0}$, 
\begin{equation}
    \Psi_{N_{A}}^{0}=
\arg\min_{\Psi
_{N_{A}}\in\Omega_{A}}\ E^{\text{WF-in-DFT}}[\Psi_{N_A};\left\{
\varphi_{i}^{\rm KS}\right\}_{N_B}]
\label{minimizerA}
\end{equation}
yields a density
$\rho_{A}^{0}$ complementary to $\rho_{B}$, i.e.%
\begin{eqnarray}
\rho_{A}^{0}=\rho_{A}[\Psi_{N_{A}}^{0}]  \label{rhoA0} \\
\rho_{A}^{0}(\mathbf{r})+\rho_{B}(\mathbf{r})=\rho_{0}(\mathbf{r})
\label{rhoPsi}%
\end{eqnarray}
Thus, upon a minimization, exact ground state density $\rho_{0}$ of the composite system is obtained, in addition to the ground state energy $E_0$, as in conventional KS-DFT calculation.  

To inspect the possible dependence of the WF-in-DFT orbital embedding functional on the unknown kinetic energy functionals, use definitions of the KS\ and HK functionals, Eqs.~(\ref{EKS}) and (\ref{HK}), respectively, and rewrite Eq.~\eqref{exact} as
\begin{align}
E^{\text{WF-in-DFT}}[\Psi_{N_{A}};\left\{  \varphi_{i}^{\rm KS}\right\}_{N_B}] &  =\left\langle \Psi_{N_{A}}|\hat{T}+\hat{V}_{ee}+\hat{V}_{ext}%
|\Psi_{N_{A}}\right\rangle +E^{\text{KS}}[\left\{  \varphi_{i}^{\rm KS}\right\}
_{N_{B}}]\nonumber\\
&  +\min_{\substack{\left\{  \varphi_{i}\right\}_{N_{A}}\rightarrow\rho_{A}\\\left\{  \varphi_{i}\right\}_{N_A}\perp\left\{  \varphi_{i}^{\rm KS}\right\}_{N_B}}} T[\left\{  \varphi_{i}\right\}_{N_A}] - T_{s}[\rho_{A}] + E_{Hxc}^{\text{nadd}}[\rho_{A},\rho_{B}]\ \ \label{exact1}
\end{align}
where it has been used that $V_{ext}[\rho_{A}+\rho_{B}]=V_{ext}[\rho_{A}]+V_{ext}[\rho_{B}]$, the energy of electrons assigned to the environment reads
\begin{equation}
E^{\text{KS}}[\left\{  \varphi_{i}^{\rm KS}\right\}_{N_B}] = \sum_{i}^{N_B} \left\langle \varphi_{i}^{\rm KS}|\hat{t}|\varphi_{i}^{\rm KS}\right\rangle
 + V_{ext}[\rho_{B}] + E_{Hxc}[\rho_{B}]
\end{equation}
and the nonadditive Hartree-exchange-correlation functional $E_{Hxc}^{\text{nadd}}$, depending on densities of the active subsystem and the environment, has been defined in the conventional way as
\begin{equation}
E_{Hxc}^{\text{nadd}}[\rho_{A},\rho_{B}]=E_{Hxc}[\rho_{A}+\rho_{B}%
]-E_{Hxc}[\rho_{A}]-E_{Hxc}[\rho_{B}]
\end{equation}
Importantly, one observes that kinetic energy functionals related to the active subsystem, the third and fourth terms in Eq.~(\ref{exact1}), do not, in general, cancel in the exact functional. In fact, for any given density $\rho_A$  and a set of orbitals 
$\left\{\varphi_i^{\rm KS}\right\}_{N_B}$, the following inequality holds (for comparison with the analogous inequality for NAKP in orbital-free embedding, FDFET, see Ref.~\citenum{wesolowski2003exact})
\begin{equation}
\Delta T_s[\rho_A,\left\{\varphi_{i}^{\rm KS}\right\}_{N_B}] = 
\min_{\substack{\left\{  \varphi_{i}\right\}  _{N_{A}}\rightarrow\rho_{A}\\\left\{  \varphi_{i}\right\}_{N_A} \perp \left\{ \varphi_{i}^{\rm KS}\right\}_{N_B}}} T[\left\{  \varphi_{i}\right\}_{N_A}]
- T_{s}[\rho_{A}] \geq 0 
\label{rel4}
\end{equation}
This kinetic-energy difference is not guaranteed to vanish even for the density $\rho_A^{0}$, see Eq.~(\ref{rhoA0}), obtained at the minimum of the exact WF-in-DFT functional. In fact, whenever the densities of subsystems $A$ and $B$ overlap, this difference is expected to be positive,
\begin{equation}
\Delta T_s[\rho_A^0, \left\{\varphi_{i}^{\rm KS} \right\}_{N_B}] > 0
\label{DeltaTs}
\end{equation}

\subsubsection{Comparison with FDET wavefunction embedding}
Let us compare the functional proposed for wavefunction embedded in the
KS-orbital-environment, Eq.~\eqref{exact1}, with its counterpart introduced in the FDET framework, i.e.\ for the orbital-free environment reading\cite{Wesolowski2008}
\begin{equation}
\Xi^{\text{FDET}}[\Psi_{N_{A}},\rho_{B}] = \left\langle \Psi_{N_A} |\hat{T}
 +\hat{V}_{ee} +\hat{V}_{ext}|\Psi_{N_A}\right\rangle 
+E^{\text{HK}}[\rho_{B}] + T_{s}^{\text{nadd}}[\rho_{A},\rho_{B}] + E_{Hxc}^{\text{nadd}}[\rho_{A},\rho_{B}]
\end{equation}
Both functionals, $\Xi^{\text{FDET}}[\Psi_{N_A},\rho_{B}]$ and
$E^{\text{WF-in-DFT}}[\Psi_{N_{A}};\left\{ \varphi_{i}^{\rm KS}\right\}_{N_B}]$ are exact but their definitions involve xc and kinetic energy density functionals whose explicit exact form is unknown. Moreover, the definition of the domain of $E^{\text{WF-in-DFT}}$, given in
Eq.~\eqref{OmA}, poses yet another difficulty and cannot be easily implemented.

Notice that, unlike in orbital-embedding WF-in-DFT, the environment density in FDET is not required to be  representable by Kohn-Sham orbitals of the composite system. However, it this condition is satisfied, it is straightforward to show that minimizers of both functionals, coincide
\begin{equation}
\forall_{\rho_{B}(\mathbf{r})=\sum_{i}^{N_B}\left\vert \varphi_i^{\rm KS}(\mathbf{r})\right\vert ^{2}}\ \ \ 
\arg\min_{\Psi_{N_{A}}}\Xi^{\text{FDET}%
}[\Psi_{N_{A}},\rho_{B}]
=\Psi_{N_{A}}^{0}=
\arg\min_{\Psi
_{N_{A}}\in\Omega_{A}}\ E^{\text{WF-in-DFT}}[\Psi_{N_{A}};\left\{
\varphi_{i}^{\rm KS}\right\}  _{N_{B}}]
\end{equation}

\subsubsection{Comparison with the projection-based-WF-in-DFT functional}
The projection-based wavefunction embedding functional
proposed by Miller et al.\cite{goodpaster2014accurate} can be written as\cite{Monino2026} 
\begin{equation}
\tilde{E}^{\text{WF-in-DFT}}[\Psi_{N_{A}};\left\{  \varphi_{i}^{\rm KS}\right\}_{N_B}] = \left\langle \Psi_{N_{A}}|\hat{T}+\hat{V}_{ee}+\hat{V}_{ext}
|\Psi_{N_A} \right\rangle +E^{\text{KS}}[\left\{ \varphi_{i}^{\rm KS}\right\}_{N_B}] + E_{Hxc}^{\text{nadd}}[\rho_{A},\rho_{B}]\ \ \label{MM}%
\end{equation}
By comparing it with the exact functional in Eq.~\eqref{exact1}, it is evident that it misses the kinetic energy term, $\Delta T_s[\rho_A,\left\{\varphi_{i}^{\rm KS}\right\}_{N_B}]$, defined in Eq.~\eqref{rel4}. 
As a result, the functional is not variational and it is not bounded from below by $E_{0}$, even if its domain were restricted to wavefunctions from the set $\Omega_{A}$. This conclusion is immediately
reached after $\tilde{E}^{\text{WF-in-DFT}}$ is
evaluated for a wavefunction$\ \Psi_{N_{A}}^{0}$ [a minimizer of the exact functional, see Eq.~(\ref{E0min})-(\ref{rhoPsi})], which results in
\begin{equation}
\tilde{E}^{\text{WF-in-DFT}}[\Psi_{N_A}^{0}; \left\{  \varphi_{i}^{\rm KS}\right\}_{N_B}] = E_{0} - \Delta T_s[\rho_A^0,\left\{\varphi_{i}^{\rm KS}\right\}  _{N_B}]
\end{equation}
Since  the kinetic energy difference $\Delta T_s$ is nonnegative, the following relation is obtained \begin{equation}
\tilde{E}^{\text{WF-in-DFT}}[\Psi_{N_{A}}^{0};\left\{  \varphi_{i}%
^{KS}\right\}  _{N_{B}}] \leq  E_{0}\ \ \ \label{under}%
\end{equation}
This proves that a minimum of a functional $\tilde{E}^{\text{WF-in-DFT}}$ is
bounded from above by the exact ground state energy $E_{0}$.
Moreover, when densities $\rho_{A}^{0}$ and $\rho_{B}$ overlap and the $\Delta T_s$ term is strictly positive, cf.\ Eq.~(\ref{DeltaTs}), minimization of the functional $\tilde
{E}^{\text{WF-in-DFT}}$ will lead to an energy strictly below $E_0$
\begin{equation}
\min_{\Psi_{N_A} \in \Omega_A}\ \tilde{E}^{\text{WF-in-DFT}}[\Psi_{N_A};\left\{  \varphi_{i}^{\rm KS}\right\}  _{N_{B}}] < E_{0}\ \ \  \label{under2}%
\end{equation}

One concludes, therefore, that even if the Miller et al.\ functional were minimized in the domain $\Omega_{A}$ and constructed with exact exchange-correlation functional, it would not lead to the exact ground state energy. Its error is negative and not smaller, in magnitude, than $\Delta T_s[\rho_A^0,\left\{\varphi_{i}^{\rm KS}\right\}_{N_B}]$
\begin{equation}
 \left\vert E_0 -\tilde{E}^{\text{WF-in-DFT}}[\Psi^0_{N_A}; \left\{ \varphi_{i}^{\rm KS}\right\}_{N_B}] \right\vert \geq \Delta T_s[\rho_A^0,\left\{\varphi_{i}^{\rm KS}\right\}_{N_B}]
\label{err}
\end{equation}
where $\Psi^0_{N_A}$ denotes a wavefunction minimizing the approximate functional.

Constraining the domain of the functional $\tilde{E}^{\text{WF-in-DFT}}$ to wavefunctions belonging to $\Omega_{A}$ is, in practice, hardly feasible. 
Although the functional derivative of $\tilde{E}^{\text{WF-in-DFT}}$ with respect to $\Psi_{N_A}$ can be readily evaluated, it remains unclear how to restrict the resulting effective Hamiltonian to the domain defined in Eq.~\eqref{OmA}.

Restricting the domain of the functional $\tilde{E}^{\text{WF-in-DFT}}$ to wavefunctions $\Psi_{N_A}$ belonging to the Hilbert space orthogonal to that spanned by the set 
$\left\{ \varphi_{i}^{\text{KS}}\right\}_{N_B}$  transforms the minimization problem into a diagonalization problem for an effective Hamiltonian. This Hamiltonian contains not only the embedding potential (arising from the functional derivative of the nonadditive Hartree-exchange-correlation energy in Eq.~\eqref{MM} with respect to $\Psi_{N_A}$) but also the appropriate projection operator, as introduced in the projection-based embedding method of Ref.~\citenum{Manby2012}.
One might speculate that violating the constraint $\Psi_{N_{A}}\in\Omega_{A}$ and instead imposing the seemingly stricter orthogonality condition on $\Psi_{N_{A}}$ leads to an
upward shift of the minimum of the WF-in-DFT energy functional relative to $E_{0}$, thereby potentially partially compensating for the error introduced in Eq.~(\ref{err}).


In summary, we have shown that the exact WF-in-DFT functional defined in
Eq.~(\ref{exact}) yields the exact ground-state energy, provided that a subset
of $N_B$ exact Kohn–Sham orbitals of the environment is available. Crucially, and contrary to a commonly held assumption, the construction of this exact functional requires, analogously
to the case of FDET, explicit knowledge of the noninteracting kinetic-energy functional as demonstrated in Eq.~(\ref{exact1}). Neglecting the unknown $\Delta T_s$ terms, as is done in the approximate WF-in-DFT functional, comes at a cost: the resulting functional is no longer variational, and its minimum is shifted downward with respect to the exact ground-state energy. 

Until now, the exchange-correlation functional used in WF-in-DFT has been assumed to be exact. In practical applications, however, one relies on approximations that inevitably introduce additional errors,  beyond those discussed above. In the remainder of this work, we focus on the strongly correlated regime, where WF-in-DFT is particularly appealing. We show that errors due to the omission of the $\Delta T_s$ terms are overshadowed by those introduced by approximate exchange-correlation functionals, which systematically underestimate the coupling between subsystems. This observation is consistent with our previous findings\cite{Beran2023,Monino2026} that identify the nonadditive xc as the primary deficiency of WF-in-DFT embedding.

\section{Computational details}
\label{section_comp_details}

\begin{figure}[!ht]
    \centering
    \subfloat[\label{h20_st}]{%
    \includegraphics[width=1\linewidth]{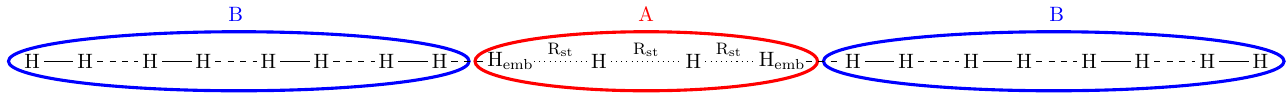}
    } \\
   \subfloat[\label{prop}]{%
    \includegraphics[width=0.6\linewidth]{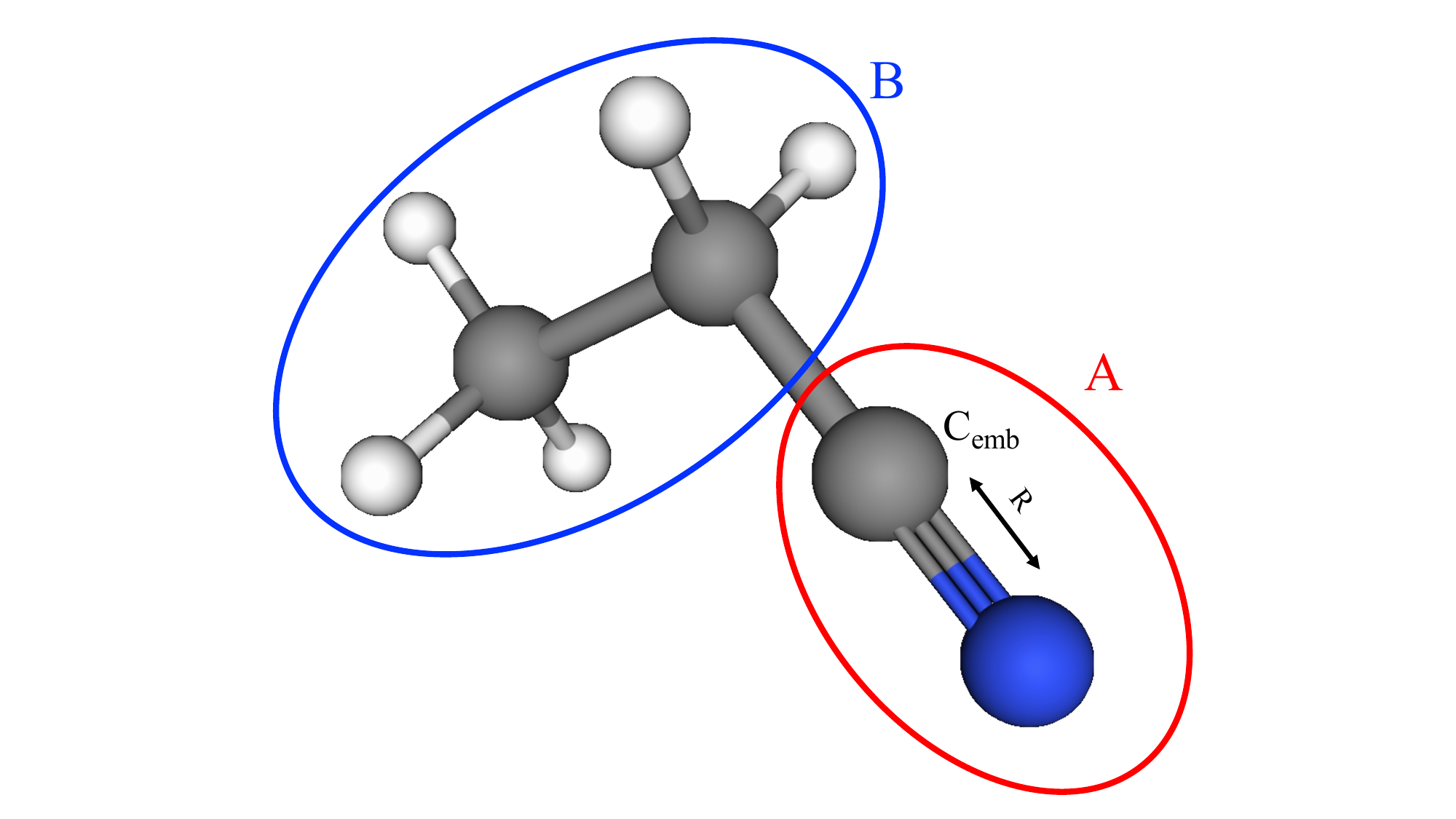}
    }
    \caption{
  Model systems investigated in this work: (a) H$_{20}$ chain with a central active fragment (A) composed of four hydrogen atoms with stretched interatomic bonds of length R$_{\text{st}}$. The environment (B) consists of hydrogen dimers with an intra-dimer bond length of 1.0~\AA. Dashed lines indicate the inter-dimer separation and the A--B distance, both equal to 1.4 \AA. H$_{\text{emb}}$ denotes hydrogen atoms that remain coupled to the environment in the dissociation limit, $\text{R}_{\text{st}} \rightarrow \infty$.  
    (b) Propionitrile (CH$_3$CH$_2$CN) molecule with the active fragment (A) defined as the $-$CN group with triple bond of the length R  stretched. C$_{\text{emb}}$ denotes a  carbon atom that remains coupled to the environment in the dissociation limit.
    }
    \label{systems}
\end{figure}

We applied the projection based WF-in-DFT of Miller et al.\cite{goodpaster2014accurate}, where we used DMRG for the WF part, to the same two prototypical molecular systems containing strongly correlated fragments that we studied in our previous work \cite{Monino2026}. As a representative example, we considered a linear chain of 20 hydrogen atoms (H$_{20}$), with the central active fragment consisting of four equally spaced hydrogen atoms, cf.\ Figure~\ref{h20_st}. To study the effect of increasing multireference character, we varied the interatomic distances within this fragment from 0.7 to 2.5 \AA. The remaining hydrogen atoms form the environment, arranged as hydrogen dimers with an intra-dimer bond length of 1.0 \AA~and an inter-dimer separation of 1.4 \AA. 
The latter distance also corresponds to the separation between the hydrogen nuclei at the edges of subsystem $A$ and the nearest hydrogen nuclei belonging to the environment $B$. As discussed below, this distance ensures strong coupling between the edge atoms of fragment~$A$ and those of fragment~$B$. To assess the effect of enlarging the active fragment, we also examined a case where the active region contains eight hydrogen atoms, the four stretched central atoms plus the two nearest hydrogen dimers (one on each side). The hydrogen chain was chosen for its one-dimensional character, as the DMRG method can provide reliable benchmark energies of near-FCI quality for the cc-pVDZ basis set \cite{Dunning1989}.

The second system we investigated is propionitrile (CH$_3$CH$_2$CN, Figure~\ref{prop}), for which we adopted the molecular geometry from our previous work on DMRG-in-DFT \cite{Beran2023}. The active subsystem was defined as the $-$CN group, and by stretching the \ce{C#N} triple bond from 0.85 to 2.5 \AA, we modulated its multireference character. In the case of propionitrile, all calculations were also performed using the cc-pVDZ basis set \cite{Dunning1989}. As a reference, we used DMRG energies within the frozen-core approximation, taken from Ref.~\citenum{Beran2023}.

For the DMRG-in-DFT calculations, the occupied orbitals, obtained from the KS-DFT computation on the entire molecule, were partitioned into active (A) and environment (B) subspaces using the SPADE procedure \cite{Claudino2019a}. A preliminary HF-in-DFT calculation was performed.
In order to maintain orthogonality between the active and environment orbitals, we employed the parameter-free approach diagonalizing the modified Fock matrix with the environment degrees of freedom projected out \cite{Khait2012}. Additionally, for DMRG-in-DFT, we employed the concentric localization (CL) technique\cite{Claudino2019b} to reduce the size of the virtual space.

All DMRG calculations, both embedded and reference, were carried out in a basis of Pipek–Mezey \cite{PipekMezey1989} split-localized molecular orbitals \cite{OlivaresAmaya2015} by means of the \textsf{MOLMPS} program \cite{Brabec2020}. Orbital ordering was optimized using the Fiedler method \cite{barcza_2011} applied to the matrix of exchange integrals \cite{OlivaresAmaya2015}, and the calculations were initialized with the CI-DEAS procedure \cite{Szalay2015, Legeza2003}. We employed the dynamical block state selection (DBSS) scheme \cite{legeza_2003a}, which adapted the bond dimension to achieve a target truncation error of $10^{-6}$. The minimum bond dimension and the fixed bond dimension used during the first warm-up sweep were both set to 500 for the embedded calculations and 1000 for the reference ones.

The DFT calculations were performed with \textsf{ORCA} package \cite{orca} using the PBE \cite{PBE1996} exchange-correlation functional. The projection-based WF-in-DFT protocol was implemented in a local version of \textsf{ORCA 5.0}.


When analyzing the performance of WF-in-DFT for a dissociating hydrogen chain, we invoke the concept of the fractional spin error (FSE)~\cite{cohen2008fractional}. The FSE diagnoses static (strong)
correlation deficiencies of approximate exchange-correlation functionals. It arises from the violation of the constancy condition, which requires the energy of a system with degenerate spin states to be invariant
with respect to the spin mixture. Such a violation leads to an artificial stabilization of spin-symmetry-broken solutions, most notably manifesting in incorrect description of covalent bond dissociation.\cite{sun2016quantum}

In this work, the FSE for a given exchange-correlation spin-density functional, $E_{xc}^{\text{spin}}[\rho_{\alpha},\rho_{\beta}]$, 
is defined as:
\begin{align}
\Delta_{xc}^{\text{FSE}}  & = E_{xc}^{\text{spin}}[\frac{\rho}{2},\frac{\rho}{2}] - E_{xc}^{\text{spin}}[\rho_{\alpha},\rho_{\beta}]\\
\rho & =\rho_{\alpha}+\rho_{\beta}
\end{align}
The first term in the definition of
$\Delta_{xc}^{\text{FSE}}$ corresponds to the equal-weight ensemble of
degenerate spin states. Typically, the FSE is positive, and for dissociating
homoatomic dimers X$_2$ it is nearly equal to one half of the dissociation
energy error obtained from spin-restricted calculations.~\cite{cohen2008fractional}

For the PBE functional\cite{PBE1996} employed in this work,
$E_{xc}^{\text{spinPBE}}[\rho/2,\rho/2]=E_{xc}^{\text{PBE}}[\rho]$, and the
fractional spin error of PBE for species X is therefore given by%
\begin{equation}
\Delta_{\text{PBE}xc}^{\text{FSE}}[\text{X}] = E_{xc}^{\text{PBE}}[\rho_{\text{X}}] - E_{xc}^{\text{spinPBE}}[\rho_{\text{X,}\alpha},\rho_{\text{X,}\beta}]
\label{FSE}
\end{equation}
where $\rho_{\text{X},\alpha}+\rho_{\text{X},\beta}=\rho_{\text{X}}$.

To assess whether eliminating the fractional spin error at the level of the exchange-correlation functional is sufficient to remedy the deficiencies observed in DMRG-in-DFT, we next turn to the pair-density functional theory (PDFT) framework~\cite{pdft}. PDFT is particularly appealing in this context because its exchange-correlation functional is, by construction, free of fractional spin error and it has been shown to be better suited to describe the xc energy of strongly correlated systems than conventional density functionals.\cite{li2014multiconfiguration,gagliardi2017multiconfiguration}

In PDFT, a pair-density functional is constructed from a parent spin-density functional using a simple ``translation'' formula that defines auxiliary spin densities $\tilde{\rho}_{\alpha}$ and $\tilde{\rho}_{\beta}$, which are then used as inputs to a conventional spin-dependent exchange-correlation functional:
\begin{equation}
\tilde{\rho}_{\alpha/\beta}(\mathbf{r})=
\frac{\rho(\mathbf{r})}{2}
\left(1\pm\sqrt{1-\frac{2\Pi(\mathbf{r})}{\rho(\mathbf{r})^{2}}}
\right)
\label{trrho}
\end{equation}
Here, $\Pi(\mathbf{r})$ denotes the on-top pair density associated with a correlated wave function, while the electron density $\rho(\mathbf{r})$ is obtained from the same wave function. In the absence of electron correlation, i.e., for a single Slater determinant,
$\Pi(\mathbf{r})=\rho(\mathbf{r})^{2}/2$, and the auxiliary spin densities
reduce to $\tilde{\rho}_{\alpha}=\tilde{\rho}_{\beta}=\rho/2$. In this limit,
the PDFT exchange-correlation energy coincides with that of the parent
density functional.

Within this framework, the PDFT exchange-correlation energy constructed from
the PBE functional is given by
\begin{equation}
E_{xc}^{\text{PDFT}}=E_{xc}^{\text{spinPBE}}[\tilde{\rho}_{\alpha}
,\tilde{\rho}_{\beta}]
\label{PDFT}
\end{equation}
and this formulation has been implemented in the \textsc{GammCor} code.~\cite{gammcor}
Because, for a given total spin $S$, both the electron density and the on-top pair density are independent of the spin projection $M_{s}$, the resulting PDFT exchange-correlation functional is, by construction, free of fractional spin error. Consistent with this property, PDFT applied on top of CASSCF wavefunctions has been shown to yield accurate potential energy curves for molecules with multiple dissociating bonds~\cite{li2014multiconfiguration}.

In the present work, we employ the PDFT exchange-correlation functional, Eq.~(\ref{PDFT}), for the evaluation of the nonadditive exchange-correlation energy within the DMRG-in-DFT embedding framework. The nonadditive xc energy is computed from the formula
\begin{equation}
E_{\text{PDFT}xc}^{\text{nadd}} = E_{xc}^{\text{PDFT}}[AB] - E_{xc}^{\text{PDFT}}[A] - E_{xc}^{\text{PBE}}[\rho_{B}]\
\end{equation}
where it has been used that the KS-DFT\ density is used for the environment $B$, thus $\tilde{\rho}_{\alpha/\beta}(\mathbf{r})=\rho(\mathbf{r})/2$, and $E_{xc}^{\text{PDFT}}[B]=E_{xc}^{\text{PBE}}[\rho_{B}]$. The PDFT xc energy for the active subsystem $A$ is evaluated
using electron density and on-top pair density corresponding to a DMRG\ wavefunction $\Psi_{N_A}$,
\begin{equation}
E_{xc}^{\text{PDFT}}[A] = E_{xc}^{\text{PDFT}}[\rho_A,\ \Pi_{\Psi_{N_A}}]
\end{equation}
Finally, the xc energy for the composite system $AB$ is computed in the PDFT framework based on a sum of densities $\rho_A$ and $\rho_{B}$, while the on-top pair density
corresponds to a wavefunction $\Psi_{N_A}\otimes\Phi_{N_B}$ which is an antisymmetrized product of $\Psi_{N_{A}}$ and the Slater determinant constructed from the KS orbitals
assigned to $B$, $\Phi_{N_{B}}$
\begin{equation}
E_{xc}^{\text{PDFT}}[AB]=E_{xc}^{\text{PDFT}}[\rho_A+\rho
_{B},\ \Pi_{\Psi_{N_{A}}\otimes\Phi_{N_{B}}}]
\end{equation}

\section{Results}
\label{section_results}

Consider the approximate WF-in-DFT\ functional presented in 
Eq.~(\ref{MM}), which can be written as%
\begin{eqnarray}
\tilde{E}^{\text{WF-in-DFT}}[\Psi_{N_A};\left\{  \varphi_{i}^{\rm KS}\right\}_{N_B}] &=& \left\langle \Psi_{N_A}|\hat{T}+\hat{V}_{ee}+\hat{V}_{ext}
|\Psi_{N_A}\right\rangle +E^{\text{KS}}[\left\{  \varphi_{i}^{\rm KS}\right\}_{N_B}] \nonumber \\
&+&\int\int\frac{\rho_{A}(\mathbf{r})\rho_{B}(\mathbf{r}^{\prime})}{\left\vert \mathbf{r-r}^{\prime}\right\vert} \, {\rm d}\mathbf{r} {\rm d}\mathbf{r'} + E_{xc}^{\text{nadd}}[\rho_A,\rho_B]\ \ \label{Approx0}%
\end{eqnarray}
where the last but one term is the nonadditive Hartree energy, and the last term denotes the nonadditive exchange-correlation (nadd xc) energy defined as
\begin{equation}
E_{xc}^{\text{nadd}}[\rho_{A},\rho_{B}]\ = E_{xc}[\rho_{A}+\rho_{B}] - E_{xc}[\rho_{A}]\ - E_{xc}[\rho_{B}] 
\label{naddXC}
\end{equation}
Both nonadditive energy terms describe interactions between the electron
densities assigned to the active fragment and the environment. More
specifically, the Hartree nonadditive energy corresponds to the Coulomb repulsion of electrons assigned to fragments $A$ and $B$, whereas $E_{xc}^{\text{nadd}}$ accounts for
quantum-mechanical effects that lead to coupling between the active subsystem
and its environment. By definition, nonadditive energy functionals vanish when
the electron densities of the fragments do not overlap.

One might expect that applying WF-in-DFT embedding
to systems for which approximate KS-DFT performs poorly would
yield accurate total energies, provided that the most challenging part of the electronic structure is included in the active subsystem $A$. This expectation is implicitly based on the
assumption that errors in the exchange-correlation energy of the composite density, $E_{xc}[\rho_{A}+\rho_{B}]$, and those of the active subsystem, $E_{xc}[\rho_{A}]$, largely cancel when forming the nonadditive exchange-correlation energy defined in Eq.~\eqref{naddXC} while all other energy terms described by DFT, i.e.\ the remaining component of the nadd xc energy $E_{xc}[\rho_{B}]$ and $E^{\text{KS}}[\left\{  \varphi_{i}^{\rm KS}\right\}_{N_B}]$ pertain to the environment, which by construction does not pose a challenge for approximate xc functionals.

As demonstrated in our previous work \cite{Monino2026}, applying WF-in-DFT (with
WF$=$DMRG) to molecules undergoing multiple covalent bond dissociation leads to potential energy surfaces that are dramatically improved compared to those obtained with KS-DFT. However, their quality remains unsatisfactory in the stretched-bond regions, where significant errors persist, even when all electrons localized on the dissociating bonds are included in subsystem $A$. This indicates that the anticipated error cancellation between the $E_{xc}[\rho_{A}+\rho_{B}]$ and $E_{xc}[\rho_{A}]$ energies is incomplete.

Using model systems, we elucidate the origin of the major error in DMRG-in-DFT. Although PBE xc functional is used here as a representative example, the problem is not specific to this functional: other semilocal and hybrid exchange–correlation functionals exhibit the same qualitative behavior.\cite{Monino2026} We also show that the PDFT functional, which is considered more suitable for strongly correlated systems, is equally deficient when employed in DMRG-in-DFT.

\subsection{Stretched hydrogen chain}
Let us first analyze a model system consisting of a linear H$_{20}$ chain (see
Figure~\ref{systems}), in which electron density of the four central hydrogen atoms is assigned to the active subsystem $A$. Consider a dissociation limit when the three central H--H covalent bonds are broken. In this limit, the electron densities decompose into sums of densities corresponding to smaller fragments, schematically represented as
\begin{align}
AB &  :\text{H}_{20}\rightarrow2\text{H}_{8}\text{H}+2\text{H}\label{ABerr}\\
A &  :\text{H}_{4}\rightarrow2\text{H}_{\text{emb}}+2\text{H}\label{Aerr}\\
B &  :2\text{H}_{8}\rightarrow2\text{H}_{8}\label{Berr}%
\end{align}
where H${_\text{emb}}$ denotes a single-electron density embedded in the H$_8$ environment and H corresponds to that of the isolated hydrogen atom.

\begin{figure}[!ht]
    \centering
    \subfloat[\label{AbsoluteH4}]{%
    \includegraphics[width=0.45\linewidth]{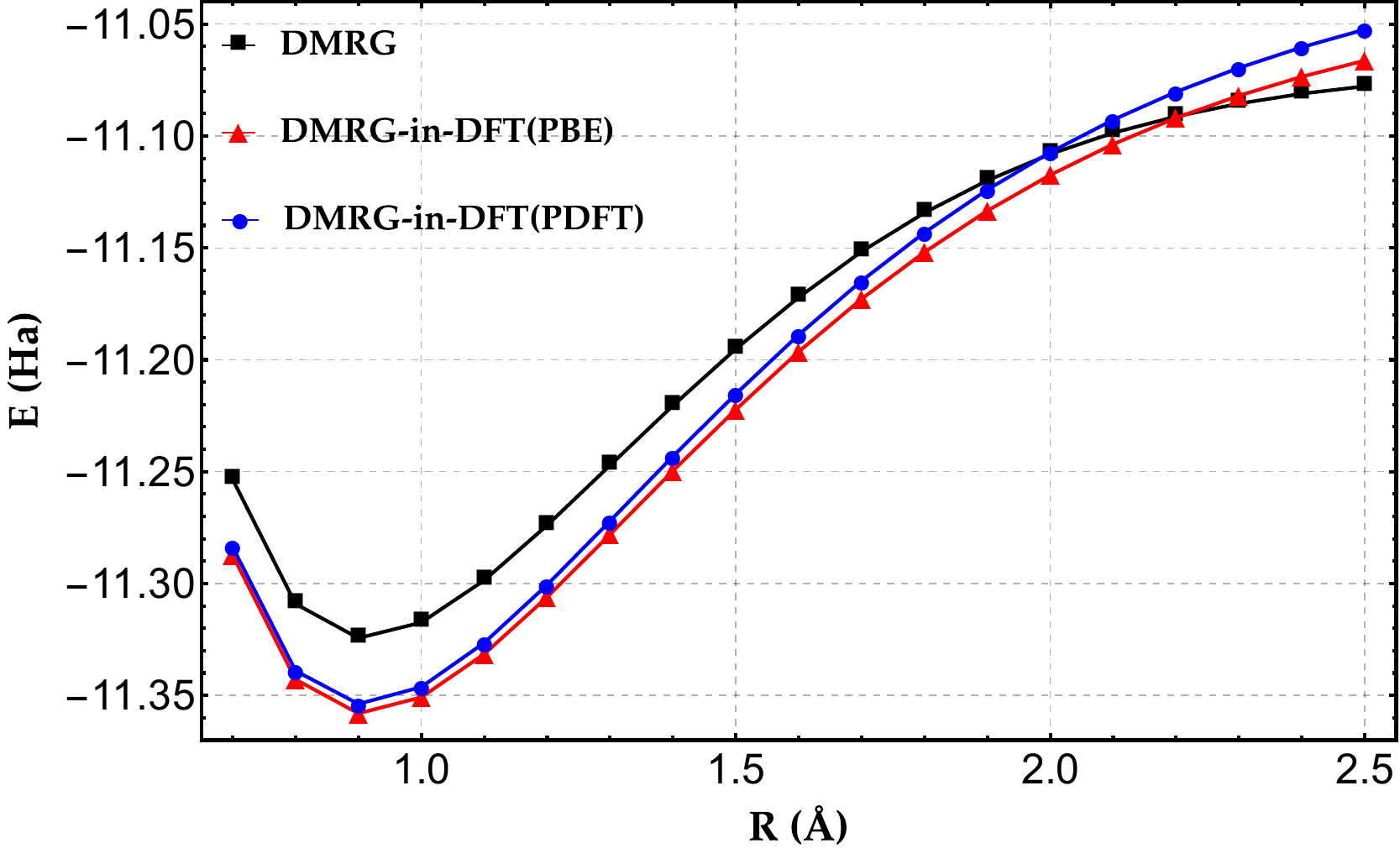}
    }
    \hskip 0.4cm
    \subfloat[\label{RelativeH4}]{%
    \includegraphics[width=0.45\linewidth]{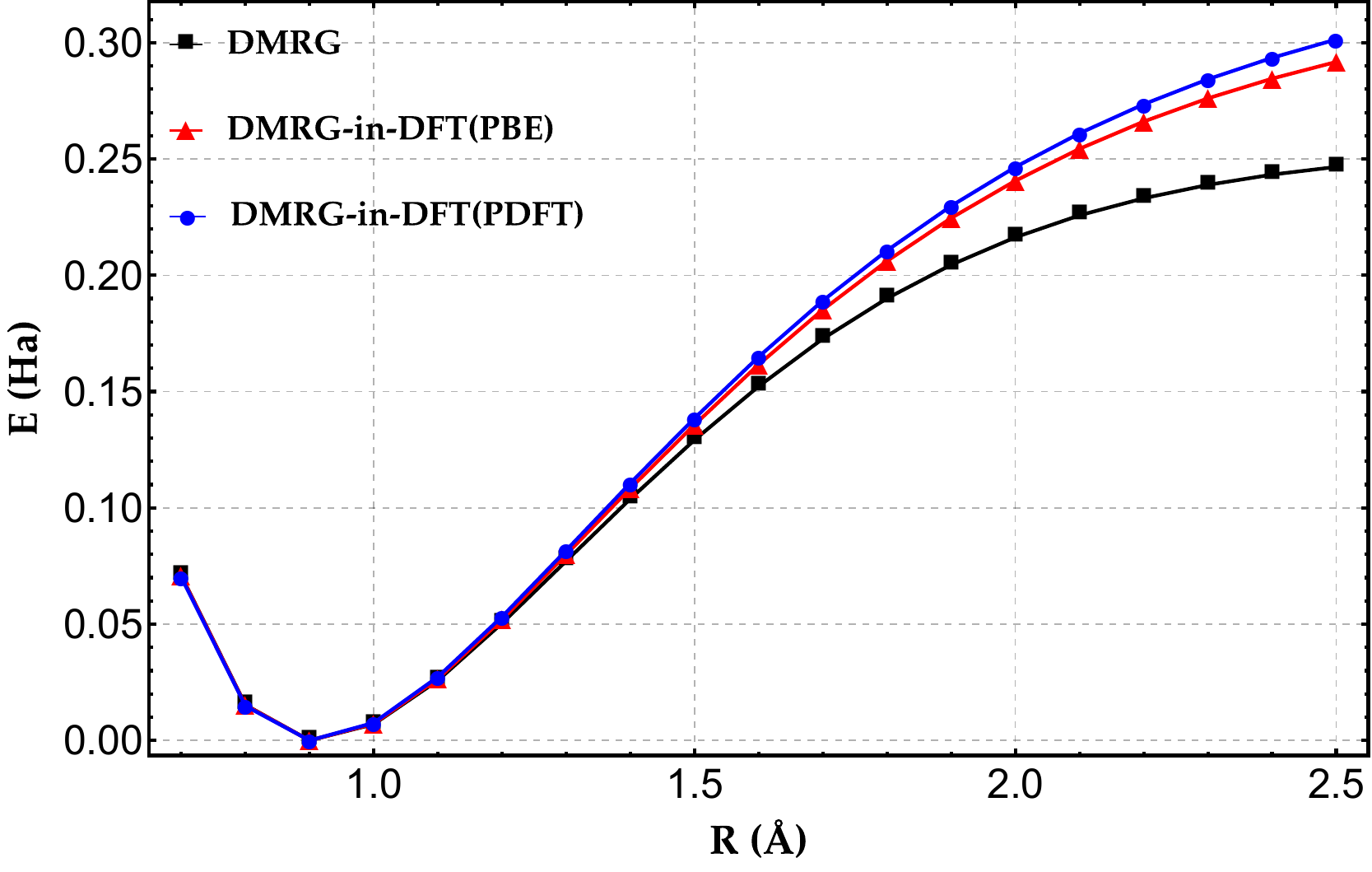}
    }
    \caption{
    Results for the H$_{20}$ chain with a four-atom active fragment using the cc-pVDZ basis set.
    (a) Absolute energies and (b) relative energies (in Ha) obtained with DMRG and DMRG-in-DFT. For the DFT part, the PBE functional was employed (see main text). Two variants are shown in which the nonadditive exchange–correlation energy (see Eq.~(\ref{Approx0})) is computed using either the PBE or the PDFT functional, denoted as DMRG-in-DFT(PBE) and DMRG-in-DFT(PDFT), respectively.  
    }
    \label{fig:h4_dmrg}
\end{figure}

Before analyzing errors in the nonadditive exchange–correlation functional, it is instructive to examine the behavior of the absolute DMRG-in-PBE energy along the dissociation coordinate, relative to the
benchmark DMRG curve (cf.\ Figure~\ref{fig:h4_dmrg}). Near the equilibrium, the DMRG-in-PBE energies lie below the benchmark values. 
This behavior can be attributed to the incompleteness of the employed atomic basis-set and to the unequal basis-set convergence of the individual energy contributions: in DMRG-in-DFT, the DFT-level terms converge more rapidly with respect to the basis set than the DMRG energy.
Upon bond stretching, however, the DMRG-in-PBE energy curve crosses the DMRG
reference and lies above it in the dissociation limit. This behavior reveals that the error of the DMRG-in-PBE energy is positive when bonds dissociate. 
Recall that the approximate treatment of the kinetic energy, specifically, setting the $\Delta T_s$ term defined in Eq.~\eqref{rel4} to zero, violates strict variationality and lowers the energy below the exact value. Since positive errors are observed in the dissociation limits, the kinetic-energy contribution must be negligible compared to the much larger error introduced by the approximate exchange–correlation functional.

Denote by $\Delta_{\text{PBE}xc}^{\text{err}}[X]$ the error in the exchange-correlation energy evaluated with the PBE functional for the electron density of fragment $X$
\begin{equation}
\Delta_{\text{PBE}xc}^{\text{err}}[X] = E_{xc}^{\text{PBE}}[\rho_X] - E_{xc}^{\text{exact}}[\rho_X]
\end{equation}
In the limit of dissociation of the H$_{20}$ chain the xc energy
error is a sum of errors of the dissociation products, cf.\ Eq.~(\ref{ABerr})%
\begin{equation}
\Delta_{\text{PBE}xc}^{\text{err}}[AB]=2\Delta_{\text{PBE}xc}^{\text{err}%
}[\text{H}_{8}\text{H}]+2\Delta_{\text{PBE}xc}^{\text{err}}[\text{H}]
\end{equation}
Errors for the considered fragments are approximately equal
to the corresponding FSE's, computed according to Eq.~\eqref{FSE}, which are positive
\begin{align}
\Delta_{\text{PBE}xc}^{\text{err}}[\text{H}_{8}\text{H}] &  =\Delta
_{\text{PBE}xc}^{\text{FSE}}[\text{H}_{8}\text{H}]\\
\Delta_{\text{PBE}xc}^{\text{err}}[\text{H}] &  =\Delta_{\text{PBE}%
xc}^{\text{FSE}}[\text{H}]
\end{align}
We verified this numerically in Ref.~\citenum{Monino2026}: removing the sum of fragment FSEs from the spin-restricted PBE energy in the dissociation limit of H$_{20}$ yields an accurate energy.
One concludes that the PBE xc energy error for the composite system is a sum of the pertinent
FSE's, namely
\begin{equation}
\Delta_{\text{PBE}xc}^{\text{err}}[AB]=2\Delta_{\text{PBE}xc}^{\text{FSE}%
}[\text{H}_{8}\text{H}]+2\Delta_{\text{PBE}xc}^{\text{FSE}}[\text{H}]
\label{errAB}
\end{equation}

Consider now the density of fragment $A$, which in the dissociation limit [see Eq.~\eqref{Aerr}] is the sum of two nonoverlapping densities of isolated hydrogen atoms and two embedded one-electron densities coupled to fragment $B$. Electron density of the H$_{\text{emb}}$ fragment, $\rho_{\text{H}_{\text{emb}}}$, differs from that of the isolated hydrogen atom, $\rho_{\text{H}}$. Although both are one-electron densities, $\rho_{\text{H}}$ is localised on a single nucleus, whereas $\rho_{\text{H}_{\text{emb}}}$ is delocalised and overlaps with the H$_{8}$ fragment. Exchange-correlation energy of a one-electron species should exactly cancel the Hartree energy. With the PBE xc functional it is not the case, and the sum of the exchange-correlation and the Hartree energies, i.e.\ the self-interaction error (SIE), is positive for H and H$_{\text{emb}}$, as shown in Table~\ref{tab:1}. 
For the isolated hydrogen atom H, the SIE (the measure of the PBE xc error) is practically equal to the FSE, i.e.\
$\Delta_{\text{PBE}xc}^{\text{err}}[\text{H}] =\text{SIE}[\text{H}] \approx \Delta_{\text{PBE}xc}^{\text{FSE}}[\text{H}]$. Thus, the xc energy error of the dissociated fragment $A$ can be written as
\begin{equation}
\Delta_{\text{PBE}xc}^{\text{err}}[A] = 2\Delta_{\text{PBE}xc}^{\text{FSE}}[\text{H}] + 2\Delta_{\text{PBE}xc}^{\text{err}}[\text{H}_{\text{emb}}]
\label{errA}
\end{equation} 

\begin{table}[h]
\centering
\caption{
The PBE exchange-correlation energy ($E_{xc}^{\text{PBE}}$), the fractional-spin error ($\Delta_{\text{PBE}xc}^{\text{FSE}}$), the PDFT exchange-correlation energy ($E_{xc}^{\text{PDFT}}$), the Hartree energy ($E_H$), and the self-interaction error (SIE), defined for one-electron densities as the sum of the PBE exchange-correlation and Hartree energies, for selected fragment densities. Energies are given in Ha.}

\label{tab:1}
\begin{tabular}{lrrrrr}
Fragment & $E_{xc}^{\text{PBE}}$ & $\Delta_{\text{PBE}xc}^{\text{FSE}}$ & $E_{xc}^{\text{PDFT}}$ & $E_{H}$ & SIE \\
\hline
H$_8$H  & -2.818 &	0.027	& -2.844		\\
H$_\text{emb}$ & -0.238	& 0.038	& -0.276	& 0.248 &	0.010 \\
H    & -0.269 &	0.043	&	& 0.313 &	0.044 \\
\end{tabular}
\end{table}

We now focus on the error of the nonadditive xc energy ($\Delta_{\text{PBE}xc}^{\text{nadd}}=\Delta^{\rm err}_{\text{PBE}xc}[AB]-\Delta^{\rm err}_{\text{PBE}xc}[A]-\Delta^{\rm err}_{\text{PBE}xc}[B]$) in the
dissociation limit. Combining Eqs.~(\ref{errAB}) and (\ref{errA}) and assuming that exchange-correlation energy for the (closed-shell and weakly correlated) environment B is accurate ($\Delta^{\rm err}_{\text{PBE}xc}[B]=0$), we obtain
\begin{equation}
\Delta_{\text{PBE}xc}^{\text{nadd}}=2\Delta_{\text{PBE}xc}^{\text{FSE}%
}[\text{H}_{8}\text{H}]-2\Delta_{\text{PBE}xc}^{\text{err}}[\text{H}%
_{\text{emb}}]>0 
\label{errPBE}
\end{equation}

The DMRG-in-PBE error derived in Eq.~\eqref{errPBE} indicates that there occurs only a partial error cancellation between $E_{xc}[\rho_A+\rho_B]$ and $E_{xc}[\rho_A]$. Namely, the exchange–correlation energy errors of the hydrogen fragments cancel exactly, but the fractional-spin error of the $AB$ fragment is not compensated by the error of fragment~$A$, which is of a different origin. A quantitative estimate of the nadd xc error in the dissociation limit can be obtained by combining the FSE of the H$_8$H fragment (27~mHa, see Table~\ref{tab:1}) with the error of H$_{\text{emb}}$, equal to the SIE (10~mHa), yielding a total of 34~mHa.

Turning to the performance of PDFT employed for the nonadditive exchange-correlation energy, Figure~\ref{fig:h4_dmrg} shows that it does not improve the DMRG-in-PBE results; in fact, it further deteriorates them. To understand this behavior, we analyze the nadd xc energy described by PDFT, denoted as $E_{\text{PDFT}xc}^{\text{nadd}}$, in the H$_{20}$ dissociation limit. Taking into account cancellation of the PDFT energy of the hydrogen atom densities, appearing in the $AB$ and $A$ fragments, see Eqs.~(\ref{ABerr})–(\ref{Berr}), one obtains
\begin{equation}
E_{\text{PDFT}xc}^{\text{nadd}}=2E^\text{PDFT}_{xc}[\text{H}_{8}\text{H}]-2E^\text{PDFT}_{xc}[\text{H}%
_{\text{emb}}]-2E^\text{PDFT}_{xc}[\text{H}_{8}]
\end{equation}
Our goal is to compare the error of the nadd xc PDFT energy, which will be denoted as $\Delta_{\text{PDFT}xc}^{\text{nadd}}$, with the error of PBE, namely with $\Delta_{\text{PBE}xc}^{\text{nadd}}$.
To enable such a comparison, first observe, see Table~\ref{tab:1},
that the PDFT exchange-correlation energy of the fragment H$_8$H is approximately equal to the PBE xc energy corrected for the fractional spin error, 
$E_{xc}^{\text{PDFT}}[$H$_8$H$]\approx E_{xc}^{\text{PBE}}[\rho_{\text{H}_{8}\text{H}}] - \Delta_{\text{PBE}xc}^{\text{FSE}}[\text{H}_{8}\text{H}]$. For the one-electron fragment H$_{\text{emb}}$, the on-top pair density is trivially zero, thus, the PDFT xc energy, recall  definitions in Eqs.~(\ref{trrho}) and (\ref{PDFT}), is just equal to the spin-PBE xc energy, $E_{xc}^{\text{spinPBE}}[\rho
,0]$. This leads to the exact equality  
\begin{equation}
    E_{xc}^{\text{PDFT}}[\text{H}_{\text{emb}}]=E_{xc}^{\text{PBE}}[\rho_{\text{H}_{\text{emb}}}] - \Delta_{\text{PBE}xc}^{\text{FSE}}[\text{H}_{\text{emb}}]
    \label{PDFTHemb}
\end{equation}
Since the environment (the H$_{8}$ fragment) is described by a single determinant, the PDFT and PBE exchange-correlation energies are identical, $E^\text{PDFT}_{xc}[\text{H}_{8}]=E^\text{PBE}_{xc}[\rho_{\text{H}_{8}}]$. Finally, the nadd xc PDFT energy 
can be written as follows
\begin{equation}
E_{\text{PDFT}xc}^{\text{nadd}}=E_{xc}^{\text{PBE}}[\rho_{\text{H}_{8}%
\text{H}}]-\Delta_{\text{PBE}xc}^{\text{FSE}}[\text{H}_{8}\text{H}%
]-2(E_{xc}^{\text{PBE}}[\rho_{\text{H}_{\text{emb}}}]-\Delta_{\text{PBE}%
xc}^{\text{FSE}}[\text{H}_{\text{emb}}])\ -2E_{xc}^{\text{PBE}}[\rho
_{\text{H}_{8}}]\label{PDFT2}%
\end{equation}

As discussed above, the PBE xc energy error for the H$_8$H fragment is
dominated by the fractional-spin error; subtracting the latter from the PBE energy (see the first two terms in the above equation) leads therefore to an accurate xc energy of this fragment.  For the H$_8$ fragment, the PBE xc energy is assumed to be accurate. It follows that the dominant contribution to the error in the nonadditive PDFT xc energy
must originate from the inaccuracy of the H$_{\text{emb}}$ xc energy. The latter should exactly cancel the corresponding Hartree energy, $E_{H}[\rho_{\text{H}_{\text{emb}}}]$, which allows us to quantify the error of PDFT xc for H$_{\text{emb}}$ as
\begin{equation}
\Delta_{\text{PDFT}xc}^{\text{err}}[\text{H}
_{\text{emb}}]\label{errPDFT}=    E^\text{PDFT}_{xc}[\text{H}
_{\text{emb}}]+E_{H}[\rho_{\text{H}_{\text{emb}}}] 
\end{equation}
By combining this equation with Eq.~(\ref{PDFTHemb}), and recalling that 
$E^\text{PBE}_{xc}[\rho_{\text{H}
_{\text{emb}}}]+E_{H}[\rho_{\text{H}_{\text{emb}}}] = \Delta_{\text{PBE}xc}^{\text{err}}[\text{H}
_{\text{emb}}]\label{errPDFT}$, we can estimate the error of PDFT nadd xc 
\begin{equation}
\Delta_{\text{PDFT}xc}^{\text{nadd}} \approx
-2\Delta_{\text{PDFT}xc}^{\text{err}}[\text{H}
_{\text{emb}}] = 
2\Delta_{\text{PBE}xc}^{\text{FSE}}[\text{H}_{\text{emb}}]-2\Delta_{\text{PBE}xc}^{\text{err}}[\text{H}%
_{\text{emb}}]\label{errPDFT}
\end{equation}
Having  identified the last term as the SIE, we estimate the PDFT nadd xc error to be 56~mHa in the dissociation limit (Table~\ref{tab:1}).

By comparing the errors of the nonadditive exchange-correlation energies
computed using PBE\ and PDFT functionals, Eqs.~\eqref{errPBE} and
\eqref{errPDFT}, it is clear that they differ in the FSE\ of H$_{\text{emb}}$ and H$_{8}$H fragments. The former is more positive than the latter, see Table~\ref{tab:1}, which leads to the relation
\begin{equation}
\Delta_{\text{PDFT}xc}^{\text{nadd}}>\Delta_{\text{PBE}xc}^{\text{nadd}}>0
\label{relation}
\end{equation}
Both functionals fail at describing the xc energy of the active H$_{\text{emb}}$ fragment, whose electron density becomes increasingly delocalized over the environment upon bond dissociation.

\subsection{Propionitrile with a dissociating CN group}
\begin{figure}
    \centering
    \subfloat[\label{AbsolutePropio}]{%
    \includegraphics[width=0.45\linewidth]{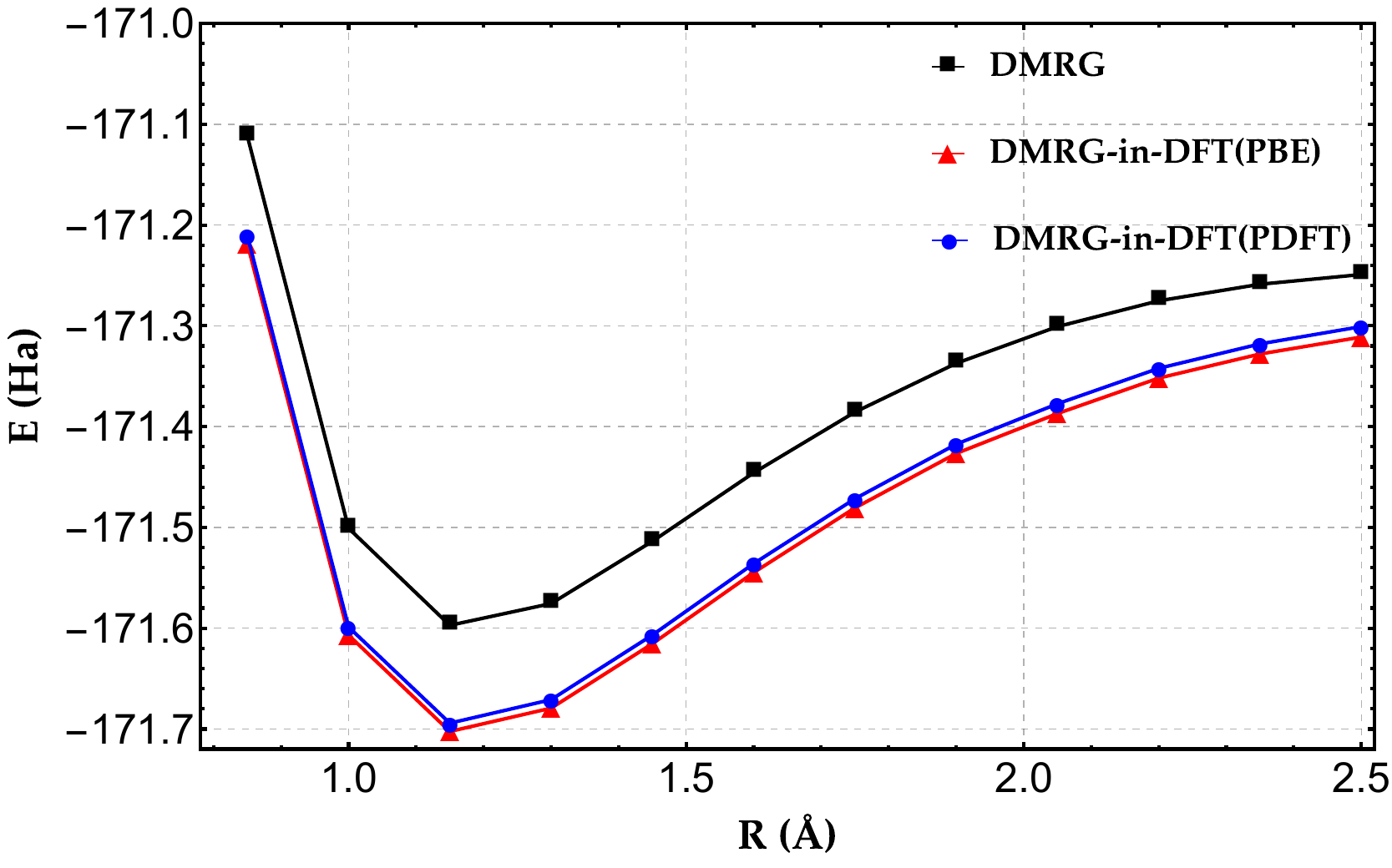}
    }
    \hskip 0.4cm
    \subfloat[\label{RelativePropio}]{%
    \includegraphics[width=0.45\linewidth]{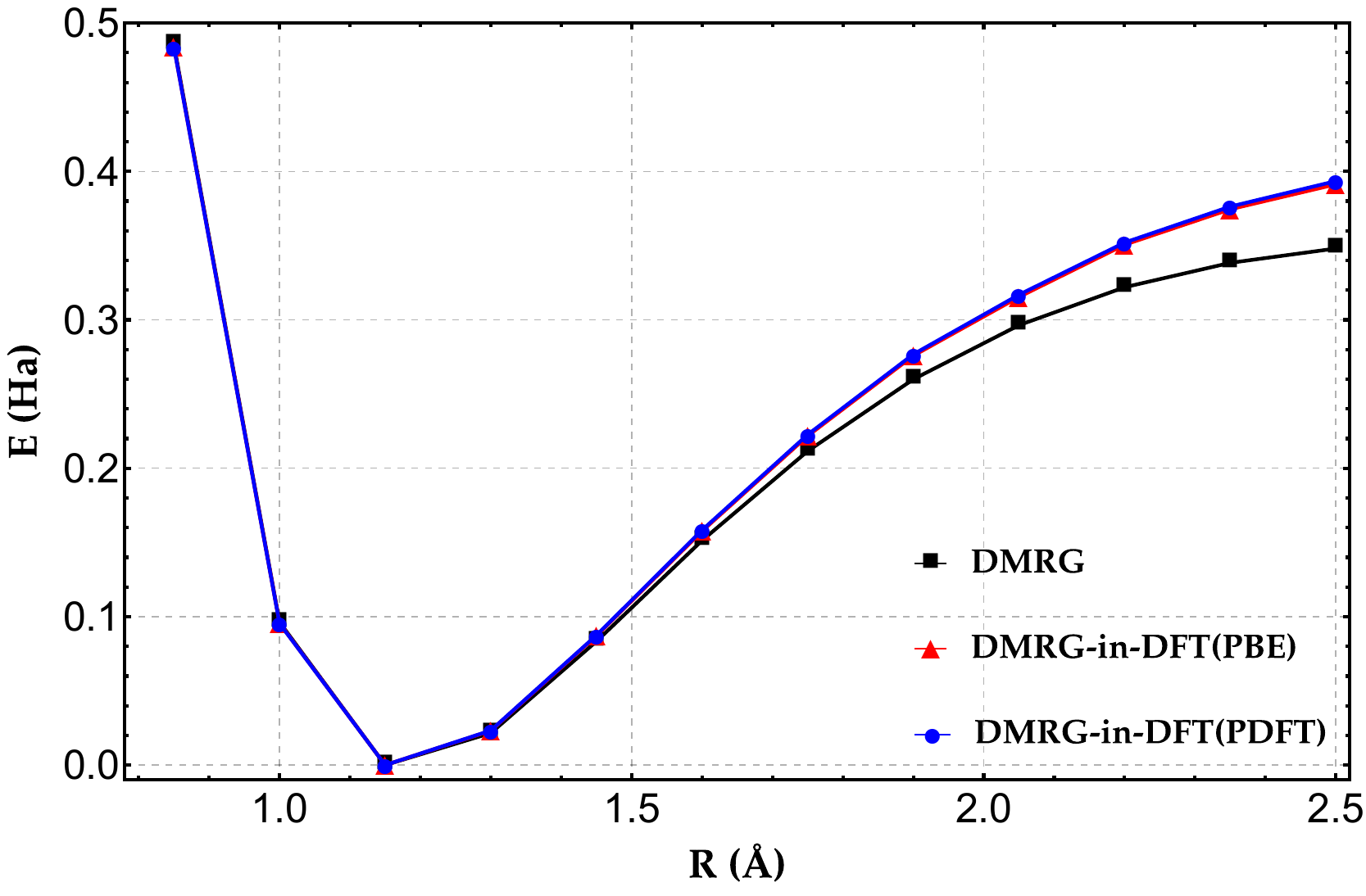}
    }
    \caption{
    Results for the propionitrile molecule using the cc-pVDZ basis set.
    (a) Absolute energies and (b) relative energies (in Ha) obtained with DMRG and DMRG-in-DFT. For the DFT part, the PBE functional was employed (see main text). Two variants are shown in which the nonadditive exchange–correlation energy (see Eq.~(\ref{Approx0})) is computed using either the PBE or the PDFT functional, denoted as DMRG-in-DFT(PBE) and DMRG-in-DFT(PDFT), respectively. 
    }
    \label{fig:propio_dmrg}
\end{figure}

For propionitrile with a dissociating CN group, DMRG-in-DFT shows the same qualitative trend as in the hydrogen chain: the error becomes positive in the stretched-bond regime (see Figure~\ref{fig:propio_dmrg}). Recall that fragment $A$ consists of a 14-electron CN group, which dissociates into a free nitrogen atom (N) and a 7-electron subfragment, denoted below as C$_{\text{emb}}$, that remains coupled to the environment. The dissociation of the  molecule and of the individual fragments can be symbolically written as
\begin{align}
AB & :\text{C}_2\text{H}_5\text{CN}\rightarrow\text{C}_2\text{H}_5\text{C}+\text{N} \\
A & :\text{CN}\rightarrow\text{C}_{\text{emb}}+\text{N} \label{cemb}\\
B & :\text{C}_2\text{H}_5\rightarrow\text{C}_2\text{H}_5
\end{align}
Assuming that the energies of the weakly correlated fragment~$B$ and the strongly correlated fragment~$A$ are accurately described by PBE and DMRG, respectively, the dominant contribution to the error of DMRG-in-DFT originates from the nonadditive exchange-correlation energy.

In the dissociation limit the xc energy of the free nitrogen-atom density cancels between the $AB$ and $A$ terms, so that the nadd xc energy reads
\begin{equation}
E_{\text{PBE}xc}^{\text{nadd}} = 
E^{\text{PBE}}_{xc}[\rho_{\text{C}_2\text{H}_5\text{C}}]
-E^{\text{PBE}}_{xc}[\rho_{\text{C}_{\text{emb}}}] - E^{{\text{PBE}}}_{xc}[\rho_{\text{C}_2\text{H}_5}]
\label{nadd_propio}
\end{equation}
Provided that the PBE xc energy predicted for the environment (\ce{C2H5}) density is accurate, the error in nadd xc is given by the difference between the PBE xc errors for the \ce{C2H5C} and \ce{C_{\text{emb}}} fragments, namely
\begin{equation}
\Delta_{\text{PBE}xc}^{\text{nadd}}=\Delta^{\text{err}}_{\text{PBE}xc}[\text{C}_{2}\text{H}_{5}\text{C}%
]-\Delta^{\text{err}}_{\text{PBE}xc}[\text{C}_{\text{emb}}]\label{err_propio}%
\end{equation}

One might suspect that the large positive error of the nonadditive energy computed with PBE originates from incomplete cancellation of the fractional-spin errors. However, Table~\ref{tab:2} shows that the FSEs of the relevant fragments, i.e., C$_2$H$_5$C and C$_\text{emb}$, agree to within 3~mHa and therefore nearly cancel in the nadd xc energy. The FSE is thus not the major source of the error.  Consistently, replacing PBE with PDFT in the evaluation of the nonadditive xc energy does not  reduce the error, as illustrated in Figure~\ref{fig:propio_dmrg}.

\begin{table}
\centering
\caption{
The PBE exchange-correlation energy ($E_{xc}^{\text{PBE}}$), the fractional-spin error ($\Delta_{\text{PBE}xc}^{\text{FSE}}$), and the PDFT exchange–correlation energy ($E_{xc}^{\text{PDFT}}$), for the products of the dissociation of the propionitrile molecule (C$_2$H$_5$) and the CN fragment (C$_{\text{emb}}$). Energies are given in Ha.}
\label{tab:2}
\begin{tabular}{lrrrrr}
Fragment & $E_{xc}^{\text{PBE}}$ & $\Delta_{\text{PBE}xc}^{\text{FSE}}$ & $E_{xc}^{\text{PDFT}}$\\ \hline
C$_2$H$_5$C    & -17.785 & 0.022 & -17.872 \\
C$_\text{emb}$ & -5.525  & 0.025 & -5.626  \\ \hline
$E_{\text{PDFT}xc}^{\text{nadd}}-E_{\text{PBE}xc}^{\text{nadd}}$ & 0.013
\end{tabular}
\end{table}

Evaluating the nonadditive xc energy in the dissociation limit [see Eq.~(\ref{nadd_propio})] using the PBE and PDFT values reported in Table~\ref{tab:2}, and noting that the exchange-correlation energies of the single-determinantal fragment~$B$ [the last term in Eq.~(\ref{nadd_propio})] are identical in PBE and PDFT,  shows that the PDFT nonadditive xc contribution is approximately 13~mHa larger (i.e., less negative) than the corresponding PBE value. Hence, the relation given in Eq.~\eqref{relation}, established for the dissociating hydrogen chain, also holds for propionitrile in the dissociation limit.


In summary, although the fractional-spin errors of PBE largely cancel in the nonadditive exchange-correlation energy for both studied systems, a significant overall error remains. PDFT, which by construction is free from the fractional-spin error, does not resolve this problem. We attribute this failure to the delocalization of the active-fragment density into the environment during bond stretching. Semilocal approximations underestimate the exchange-correlation coupling between subsystems, leading to overstabilization of fragment $A$ relative to the composite system $AB$, i.e., the nonadditive xc energy is insufficiently negative.

\begin{figure}[!ht]
    \centering
    \subfloat[\label{AbsoluteH8}]{%
    \includegraphics[width=0.45\linewidth]{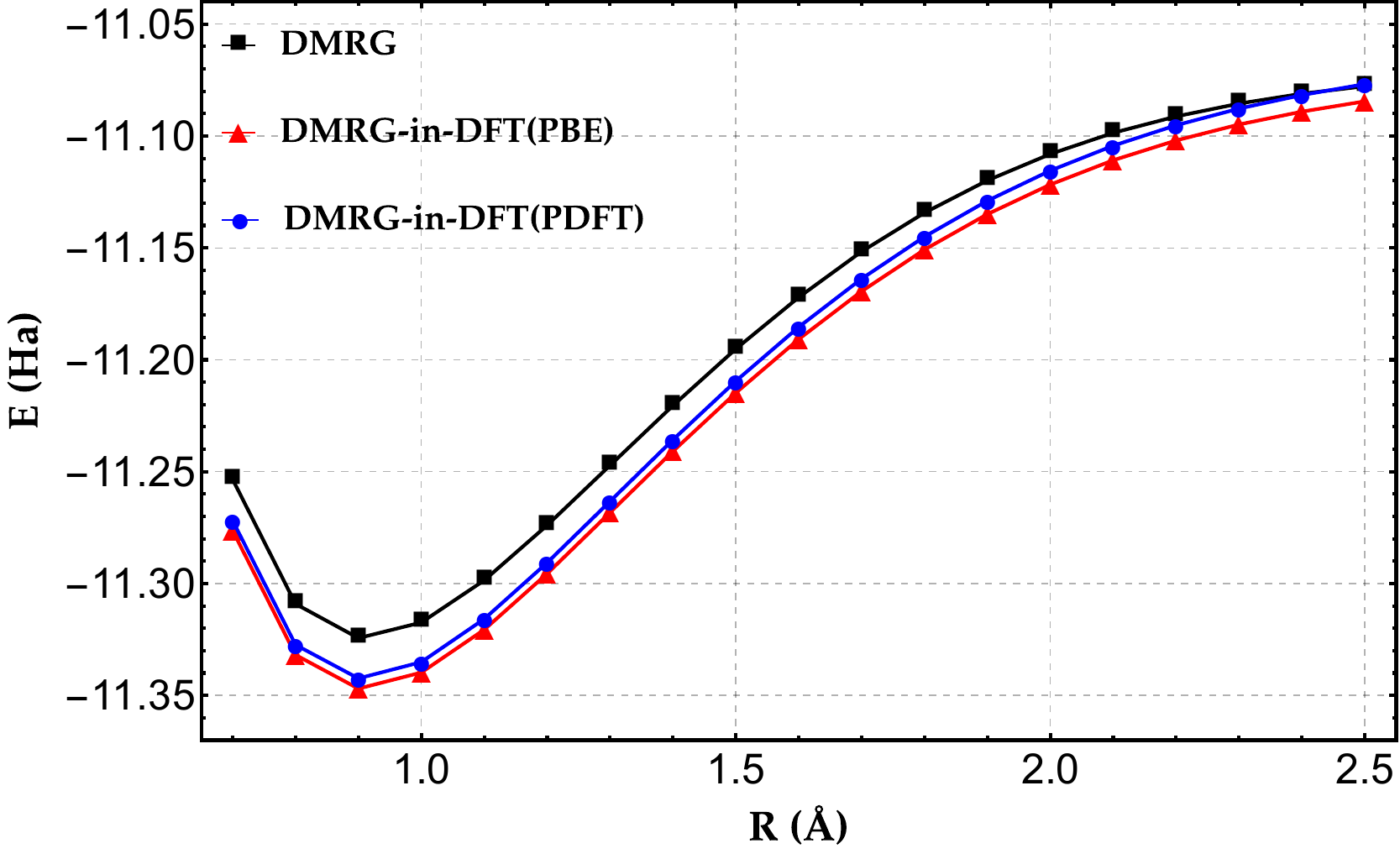}
    }
    \hskip 0.4cm
    \subfloat[\label{RelativeH8}]{%
    \includegraphics[width=0.45\linewidth]{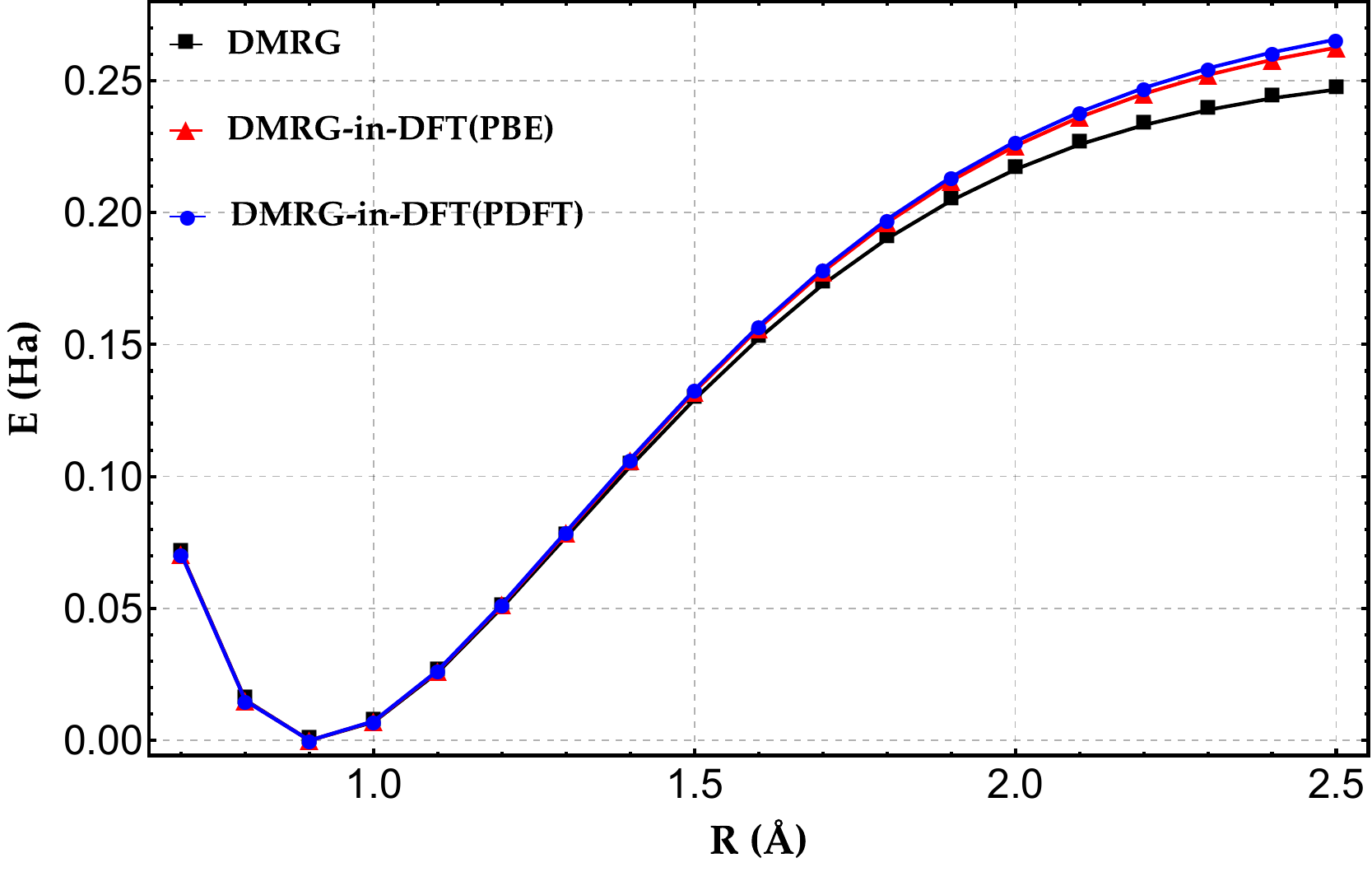}
    }
    \caption{
    Results for the H$_{20}$ chain with a 8-atom active fragment using the cc-pVDZ basis set. 
    (a) Absolute energies and (b) relative energies (in Ha) obtained with DMRG and DMRG-in-DFT. For the DFT part, the PBE functional was employed (see main text). Two variants are shown in which the nonadditive exchange–correlation energy (see Eq.~(\ref{Approx0})) is computed using either the PBE or the PDFT functional, denoted as DMRG-in-DFT(PBE) and DMRG-in-DFT(PDFT), respectively. }
    
    \label{fig:h8_dmrg}
\end{figure}
\begin{figure}[!ht]
    \centering
   
    \includegraphics[width=0.8\linewidth]{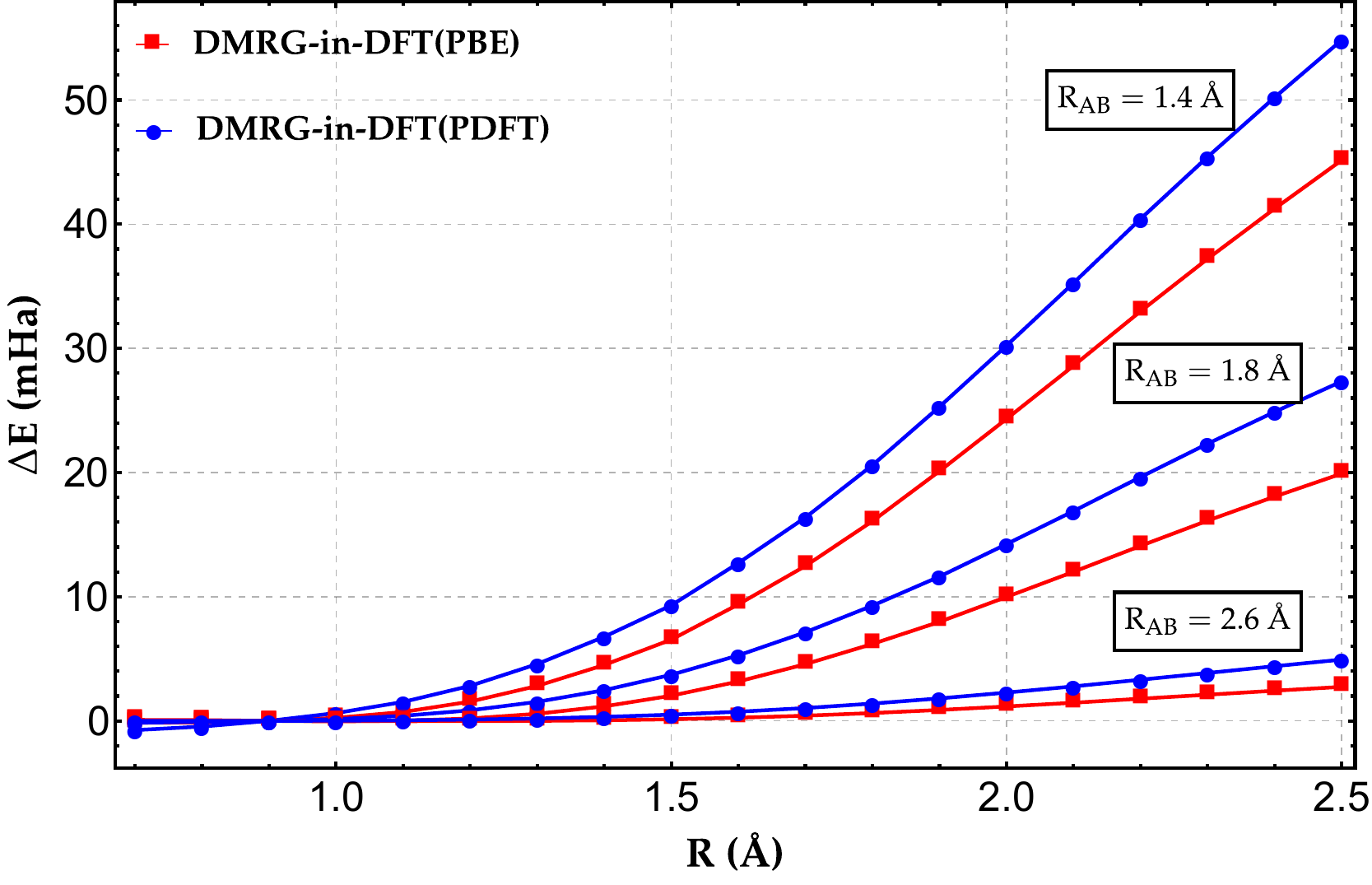}
    
    \caption{
  Relative error energies (in mHa) for the H$_{20}$ chain with a 4-atom active fragment obtained with DMRG and DMRG-in-DFT using the cc-pVDZ basis set.  Results are shown for three separations between the active fragment (A) and the environment (B): R$_{\text{AB}}$ = 1.4 \AA, R$_{\text{AB}}$ = 1.8 \AA~and R$_{\text{AB}}$ = 2.6 \AA. For each separation, two variants are presented in which the nonadditive exchange-correlation energy [see Eq.~(\ref{Approx0})] is computed using either the PBE or the PDFT functional, denoted as DMRG-in-DFT(PBE) and DMRG-in-DFT(PDFT), respectively. }
    \label{fig:R_AB}
\end{figure}

Figure~\ref{fig:h8_dmrg} corroborates this interpretation. It illustrates that increasing the number of H atoms included in fragment $A$ of the H$_{20}$ chain from 4 to 8  reduces the error. Enlarging fragment $A$ suppresses variations in the overlap of densities of the fragments, and consequently in the nonadditive exchange–correlation energy, diminishing relative energy errors. 
A similar effect is observed upon increasing the distance between the active system and the environment, see Figure~\ref{fig:R_AB}. 
Since this reduces both the density overlap and the xc coupling between the subsystems, the corresponding decrease in error is expected. Indeed, as it can be seen, for both PBE and PDFT xc functionals used in nadd xc, the relative error drops by around 25 mHa for the stretched-bond system, when the distance between edge nuclei of the subsystems is increased by 0.4 \AA.

\section{Summary and Conclusions}
\label{section_conclusions}
The projection-based WF-in-DFT embedding approach was originally proposed to  avoid explicit approximations to the kinetic energy density functional, which is a major practical limitation of frozen-density embedding theory. In this work, we formulated an exact WF-in-DFT embedding framework in which the embedded wave function is defined in an environment represented by a selected subset of Kohn–Sham orbitals of the full system.
By comparing this exact formulation with the commonly used projection-based WF-in-DFT functional introduced by Miller and coworkers \cite{Manby2012}, we demonstrated that the latter cannot reproduce the exact ground-state energy, even in the hypothetical limit of an exact exchange-correlation functional. Specifically, projection-based embedding neglects the nonnegative kinetic-energy contribution, $\Delta T_s$ [cf.\ Eq.~(\ref{rel4})], present in the exact functional. As a consequence, the method is intrinsically non-variational and yields total energies that approach the exact ground-state energy from below.


In this work, we further analyzed the main source of inaccuracy in projection-based DMRG-in-DFT employing a semilocal xc functional (PBE in the present study). We find that the missing kinetic-energy contribution plays only a minor role, whereas the dominant error originates from the nonadditive exchange-correlation energy. Using the dissociating hydrogen chain as an example, we show that errors in the nonadditive exchange-correlation energies of fragments AB and A do not cancel. Specifically, the error in AB is driven by the FSE, while the residual error in fragment A stems from the SIE associated with the delocalization of A’s density over fragment B. An analogous lack of error cancellation is observed in the CN dissociation limit of propionitrile.


By incorporating the wave-function on-top pair density, the recently introduced PDFT exchange-correlation functionals are capable of describing covalent bond breaking. An additional advantage is that, by construction, they are free from the FSE. Motivated by these features, we applied PDFT to evaluate the nonadditive exchange-correlation energy of our model systems. Contrary to expectations, PDFT does not improve upon PBE and, in fact, performs slightly worse. The reason is that PDFT also suffers from the SIE, and there is no error cancellation in the nonadditive exchange-correlation energy.

In conclusion, the dominant source of error of DMRG-in-DFT lies in the nonadditive exchange-correlation functional, making its improvement essential for achieving higher accuracy. As demonstrated in our previous work,\cite{DMRG-AC0} a proper noninteracting exchange-correlation correction can substantially reduce these errors in model systems.  These results indicate that, despite intrinsic limitations, DMRG-in-DFT can attain high accuracy when the nonadditive xc functional is properly treated.

\section*{Acknowledgments}

This work was supported by the Czech Science Foundation (Grant No. 25-18486S); the National Science Center of Poland (Grant No. 2021/43/I/ST4/02250), and the Center for Scalable and Predictive methods for Excitation and Correlated phenomena (SPEC), which is funded by the U.S. Department of Energy (DOE), Office of Science, Office of Basic Energy Sciences, the Division of Chemical Sciences, Geosciences, and Biosciences.

\bibliography{references}

\end{document}